\documentclass[twocolumn]{custom}
\pdfoutput=1
\usepackage{sidecap}
\usepackage{graphicx}
\usepackage{hyperref}
\usepackage{longtable}
\usepackage{multirow}
\usepackage{url}
\usepackage{rotating}

\def\apj{ApJ}
\def\aap{A\&A}
\def\aapr{A\&A~Rev.}
\def\mnras{MNRAS}
\def\apjl{APJL} 
\def\pasp{PASP} 
\def\procspie{procspie}
\usepackage{longtable}

\begin{document}
	\noindent
	\textsf{Astronomy Reports, 2018, Vol. 62, No. 12, pp. 917 -- 925.}\\
	
	\title{Prospects for observing strongly lensed supernovae behind Hubble Frontier Fields galaxy clusters with the James Webb Space Telescope}

	\author{\firstname{T.}~\surname{Petrushevska}}
	\affil{Centre for Astrophysics and Cosmology, University of Nova Gorica, Vipavska 11c, 5270 Ajdov\v{s}\u{c}ina, Slovenia}
	\email{tanja.petrushevska@ung.si}
		
		\author{\firstname{T.}~\surname{Okamura}}
	\affil{Department of Astronomy, University of Tokyo, Tokyo, Japan}
	
	\author{\firstname{R.}~\surname{Kawamata}}
	\affil{Department of Astronomy, University of Tokyo, Tokyo, Japan}
	
		\author{\firstname{L.}~\surname{Hangard}}
	\affil{Oskar Klein Centre, Department of Physics, Stockholm University, Stockholm, Sweden}
	
		\author{\firstname{A.}~\surname{Goobar}}
	\affil{Oskar Klein Centre, Department of Physics, Stockholm University, Stockholm, Sweden}
	
		\author{\firstname{G.}~\surname{Mahler}}
	\affil{University of Michigan, Department of Astronomy, Ann Arbor, USA}

\begin{abstract}{Measuring time delays from strongly lensed supernovae (SNe) is emerging as a novel and independent tool for estimating the Hubble constant (\emph{H$_0$}). This is very important given the recent discord in the value of \emph{H$_0$} from two methods  that probe different distance ranges.  The success of this technique will rely of our ability to discover strongly lensed SNe with measurable time delays. Here, we present the magnifications and the time delays for the multiply-imaged galaxies behind the Hubble Frontier Fields (HFF) galaxy clusters, by using recently published lensing models. Continuing on our previous work done for Abell 1689  (A1689) and Abell 370, we also show the prospects of observing strongly lensed SNe behind the HFF clusters with the upcoming James Webb Space Telescope (JWST). With four 1-hour visits in one year, the summed expectations of all six HFF clusters are $\sim0.5$ core-collapse (CC)~SNe and $\sim0.06$ Type Ia SNe (SNe~Ia) in F115W band, while with F150W the expectations are higher,  $\sim0.9$ CC~SNe and $\sim0.06$ SNe~Ia.  These estimates match those expected by only surveying A1689, proving that the performance of A1689 as gravitational telescope is superior. In the five HFF clusters presented here, we find that F150W will be able to detect SNe Ia (SNe IIP) exploding in 93 (80) pairs multiply-imaged galaxies with time delays of less than 5 years. 
}\end{abstract}

\published{December 28, 2018}

\section{Introduction}
Clusters of galaxies are gravitational lenses which magnify the light of objects behind them. In the line of sight of gravitational lenses, multiple images of background lensed supernovae (SNe) can be observed, and due to the variable nature of the objects, the difference between the arrival times of the images can be measured. Since the images have taken different paths through space before reaching us, the time-differences are sensitive to the expansion rate of the universe. 
\citet{1964MNRAS.128..307R} proposed to measure the value of the Hubble constant (\emph{H$_0$}) from the time delays of multiply-imaged SNe (for a review of strong lensing gravitational time delays as a tool for cosmography, see  \citealt{2016A&ARv..24...11T}). With the first discovery of multiply-imaged SN in 2014 \citep{2015Sci...347.1123K}, the feasibility of measuring the value of the \emph{H$_0$} with this method became possible \citep{2018ApJ...853L..31V,2018ApJ...860...94G}. This is particularly interesting since there is a recent discrepancy of \emph{H$_0$} value of at least $\approx3\sigma$ from two independent cosmological probes (the cosmic microwave background \citep{2016A&A...594A..13P} and local distant ladders \citep{2016ApJ...826...56R, 2018ApJ...855..136R}.  

After the time delays of strongly lensed transient objects have been measured, extracting cosmological parameters requires knowledge of the lensing potential (e.g. \citealt{2003A&A...405..859G,2017MNRAS.468.2590S}). Notably, gravitational lens models suffer from the so-called mass-sheet degeneracy  \citep{2003MNRAS.338L..25O}, so very different lens models will still be able to predict similar time delays of the multiply-imaged background source.  However, thanks to the homogeneous nature of Type Ia SNe (SNe~Ia) lightcurves, the observation of strongly lensed SNe~Ia circumvents this problem because the magnification can be inferred directly. Conversely, if the background cosmology is assumed to be known, even a single magnified SNe~Ia could be used to put constraints on the lensing potential  \citep{2014ApJ...786....9P,2014MNRAS.440.2742N, 2015ApJ...811...70R}.  
The magnification boost provided by the cluster lenses enhances the possibility of observing SNe at unprecedented distances, which enables to study their properties and their possible evolution with redshift that otherwise would be undetectable \citep{2017A&A...603A.136P}. Furthermore, discoveries of high redshift SNe can be used to  pinpoint the SN rates at very high redshifts ($z >2$) where they are poorly measured \citep{2000MNRAS.319..549S,2003A&A...405..859G,2016A&A...594A..54P}.

 For the purpose of using them as gravitational lenses, here we focus on the galaxy clusters from the  {\it Hubble Frontier Fields} (HFF)  \citep{2017ApJ...837...97L} with the Hubble Space Telescope (HST). The HFF program consisted of imaging six clusters - Abell 2744, MACS J0416.1-2403, MACS J0717.5+3745, MACS J1149.5+2223, Abell S1063 and Abell 370 (their short names are listed in Table~\ref{HFF_clusters}). Using these deep observations and complemented by many follow-up multi-wavelength imaging and spectroscopy, these massive clusters have been extensively modelled  and dozens of multiply-imaged galaxies  behind them have been confirmed (e.g. \citealt{2015MNRAS.446.4132J,2015A&A...574A..11K,2016A&A...588A..99L,2017MNRAS.469.3946L, 2018MNRAS.473..663M, 2018ApJ...855....4K}). New strong lens mass models of AS1063, as well as updated mass models of A2744 and MACSJ0416 have been published in (\citealt{2018ApJ...855....4K}, K18 hereafter) and (\citealt{2018MNRAS.473..663M}, M18). Those models exploit the novel dataset of very deep HST photometry and VLT/MUSE integral field spectroscopy.  The large number of spectroscopic redshifts improves the ability of the models to predict accurate magnification and time delays. Therefore, strengthens the reliability of those model predictions and the outcome of this work. The models considered in this study are computed using two different parametric approaches. K18 is using the GLAFIC \citep{2001astro.ph..2340K}, whereas M18 uses Lenstool \citep{2007NJPh....9..447J}. The choice of the type and the number of potential to model the cluster is the main differences between the two models (see \citealt{2017MNRAS.472.3177M} for more details).

In our previous work (\citealt{2016A&A...594A..54P,2018A&A...614A.103P}, hereafter P16 and P18), we used the galaxy clusters Abell 1689 (A1689) and A370  as gravitational telescopes to search for strongly lensed SNe.  There, we showed the expectations of search campaigns that can be conducted with future facilities. The HST successor, James Webb Space Telescope (JWST; \citealt{2018arXiv180506941K})  is to be launched in 2021 and will carry the NIRCam instrument aboard with several near-infrared filters with the possibility of reaching unprecedented sensitivity. Since most of the light of nearby SNe is in the optical bands, these near-infrared filters are optimal for finding high-$z$ SNe, as their light is redshifted to the longer wavelengths. The aim of this article is to complement the previous work by extending the simulations to the remaining HFF clusters to determine the prospects  for observing lensed SNe with the JWST.

\section{Prospects  for observing lensed supernovae behind the HFF galaxy clusters with the JWST}
\begin{table}
	\begin{center}

		\caption{	\small{Hubble Frontier Fields galaxy clusters considered in this work\label{HFF_clusters}. Only the background systems with spectroscopic redshift and those bellow $z<4.0$ are selected. The number of galaxies behind the cluster is given in column 3 and the number of multiple images of these galaxies in column 4. The highest redshift of these galaxies is given in column 5.}}
		\begin{tabular}{lcccc}
			
			\hline
	{\small Cluster }& 	{\small Short name} & 	{\small $N_{{\scriptsize sys}}$ }& 	{\small $N_{images}$} & 	{\small $z_{max}$ }\\ 
	\hline
	{\small 	Abell 2744 }& 	{\small A2744 }& {\small 12 }& {\small 40} & {\small 3.98 }\\ 
		{\small Abell S1063 }& 	{\small AS1063 }&{\small  14 }& {\small 42} &{\small  3.71}\\ 
	{\small 	MACS J1149.5+2223 }&	{\small MACSJ1149 }& {\small 8} &{\small  24} & {\small 3.70} \\ 
		{\small MACS J0416.1-2403 }&	{\small MACSJ0416} & {\small 23 }&{\small  65 }& {\small 3.87} \\ 
		{\small MACS J0717.5+3745 }&	{\small MACSJ0717} & {\small 6 }& {\small 20 }& {\small 2.96 }\\
			\hline
		\end{tabular}
	\end{center}
\end{table}
In the present study, we use the same procedures as in P16 and P18 for simulating surveys that can be performed with the JWST.  We start by considering the multiply-imaged galaxies behind the five HFF galaxy clusters with names and properties are listed in Table~\ref{HFF_clusters}. To avoid additional source of error, we select the background systems with spectroscopic redshift and those bellow $z<4.0$, so that redshifting to the observers frame and cross-filter k-corrections \citep{1996PASP..108..190K} are meaningful. The total number of systems that satisfy these criteria is 64 which have 183 multiple images. We assume 4 visits to each galaxy cluster for easy comparison with the previous work with A1689 and A370 (P16, P18). 

To obtain the expected SN rates for the systems behind the clusters, we estimated the specific star formation rates (SFR) and the stellar masses of the galaxies from  multi-band HST photometry. The deep HST photometry of the clusters were obtained during the HFF program \citep{2017ApJ...837...97L}, using three bands of the Advanced Camera for Surveys (ACS) and four bands of the Wide Field Camera 3 (WFC3) and is publicly available\footnote{http://www.stsci.edu/hst/campaigns/frontier-fields/}.  As previously, the SFR and stellar masses were fitted with the help of the software package {{FAST}}\footnote{http://w.astro.berkeley.edu/~mariska/FAST.html} \citep{2009ApJ...700..221K}. The brightness of each multiply-imaged galaxy was corrected for the magnification which we obtained from the lensing models in K16 and K18 and has uncertainty of the the order of $\sim10-15\%$. In the case of A2744, we compared the magnifications from K18 with those from M18 and found good agreement. After the SFR and the stellar masses of each galaxy is estimated, the expected SN rate can be inferred. For core-collapse (CC)~SN rate in units of $yr^{-1}$, we used the relation:
\begin{equation}
R_{\rm CC} = k_8^{50} \cdot \rm SFR,
\end{equation} 
because the progenitors of CC~SNe are short-lived stars, so it is expected that they trace ongoing SFR. The SFR is given in in units $\rm M_\odot$yr$^{-1}$, and $k_8^{50}$ is the scale factor, and is in P16,  we used $k_8^{50}=0.007 \,\rm M_\odot^{-1}$.  The SN~Ia rates were obtained from the  \citet{2005ApJ...629L..85S} simple model
\begin{equation}
R_{Ia} = A \cdot {\rm SFR} + B \cdot M_*,
\end{equation} 
where $A$ and $B$ are constants, for which we used literature values from \citet{2012ApJ...755...61S}, while the stellar mass $\rm M_*$ of the individual galaxies was taken from the FAST best fits. The magnifications of the galaxy images together with the SN rates of the unique galaxy systems are  listed in Table~\ref{table:multi}.	

Then, the expected number of SNe in each galaxy is obtained  by multiplying the SN rate and the control time. The control time is different for each SN type and it is the time that the SN light curve is above the detection threshold, thus is a function of the SN light curve, absolute intrinsic SN brightness, detection efficiency, extinction by dust and the lensing magnification. In Table~\ref{table:multi}, the galaxies that have positive control times in F150W band for all SN types are shown. The total CC control time was obtained by weighting the contribution from the various CC~SN subtypes with their fractions and then summed.  We obtain the final number of SNe by summing over the individual estimates from each multiple image in the system. The results are shown is Table~\ref{table:JWST} together with the previous estimates for A1689 and A370 from P18 for comparison. For more details regarding the estimates presented in this section, we refer to our previous work \citep{2018A&A...614A.103P,2016A&A...594A..54P,Second}. 

\begin{table}
	\caption{{\small Expectations for lensed SNe in the multiply-imaged galaxies behind the HFF galaxy clusters observed with JWST/NIRCam-like filters in one year.  The depth of the survey in both filters is assumed to be 27.5 mag, which is the limiting $5\sigma$ depth for a 1 hour exposure \citep{2012SPIE.8442E..2NB}. The separation between two epochs is set to 30 days and we assume 4 visits in one year. The errors in the N$_{\rm CC}$ and N$_{\rm Ia}$ originate from the propagated uncertainty in the SFR. 	$(a)$ The number of expected SNe in the background galaxies with resolved multiple images and secure redshifts (see Table~\ref{table:multi}). 	$(b)$ The maximum redshift of the expected SNe Ia. $(c)$ Number of galaxies that can host observable SNe Ia.}}
\label{table:JWST} \centering \begin{tabular}{lcccc} \hline  Cluster   & N$_{\rm CC}^a$ & N$_{\rm Ia}^a$& $z_{max}^b$ & N$_{gal }^c$  \\
		&/yr &/yr & &\\
		\hline	
		& F115W  && &  \\ 
			\hline
A2744 &$ 0.02 ( 0.01 ) $&$ 0.0012 ( 0.0003 )$& 3.05 & 21  \\  
AS1063 &$ 0.11 ( 0.05 ) $&$ 0.007 ( 0.003 )$& 3.12 & 36  \\  
MACSJ1149 &$ 0.025 ( 0.005 ) $&$ 0.005 ( 0.001 )$& 3.70 & 23  \\  
MACSJ0416 &$ 0.19 ( 0.06 ) $&$ 0.014 ( 0.004 )$& 3.29 & 49  \\  
MACSJ0717 &$ 0.05 ( 0.02 ) $&$ 0.007 ( 0.005 )$& 2.96 & 20  \\  
A370 &   $0.11 (0.04)$ & $0.02 (0.01)$ & 3.77 & 47 \\ 
A1689  & $0.7 (0.3)$ & $0.13(0.06)$ & 3.05  &66 \\ 
		
		\hline
& F150W  && &  \\ 
		\hline
A2744 &$ 0.06 ( 0.04 ) $&$ 0.006 ( 0.004 )$& 3.98 & 40  \\  
AS1063 &$ 0.12 ( 0.06 ) $&$ 0.008 ( 0.004 )$& 3.61 & 42  \\  
MACSJ1149 &$ 0.08 ( 0.02 ) $&$ 0.005 ( 0.001 )$& 3.7 & 24  \\  
MACSJ0416 &$ 0.24 ( 0.07 ) $&$ 0.016 ( 0.005 )$& 3.87 & 67  \\  
MACSJ0717 &$ 0.12 ( 0.07 ) $&$ 0.007 ( 0.004 )$& 2.96 & 20  \\  
A370 & $0.3(0.1)$ & $0.02 (0.01)$ &3.77 & 47  \\ 
A1689  & $1.0(0.5) $ & $ 0.14(0.07)$ & 3.05 & 66 \\ 

		\hline
	\end{tabular}

\end{table}

The total number of expected SNe of the five HFF clusters is $\sim0.4$ CC~SNe and $\sim0.04$ SNe~Ia in F115W band, while in F150W the expectations are higher,  $\sim0.6$ CC~SNe and $\sim0.04$ SNe~Ia. Thus, the summed expectations of all six HFF clusters are comparable to what is expected by only monitoring A1689 as shown in Table~\ref{table:JWST}. We note that these are a lower limits, since we have only considered the galaxies with spectroscopic redshift. With ongoing spectroscopic campaings with instruments such as VLT/MUSE (e.g. \citealt{2016A&A...590A..14B,2017MNRAS.469.3946L,2018MNRAS.473..663M}), more systems behind galaxy clusters will be confirmed.

\section{Summary and conclusions}
The recent discovery of the first multiply-lensed SNe, SN Refsdal and iPTF16geu \citep{2017Sci...356..291G}, opened a new path to measure the expansion history of the Universe through monitoring time delays of strongly lensed SNe. Measuring time delays of strongly lensed SNe will provide independent  and precise estimate of the Hubble constant, which will allow to shed light on the current tension regarding its value. 

In the present study, we focused on SNe lensed by a galaxy cluster lenses, unlike other recent studies that make predictions for SNe being lensed by galaxies (e.g. \citealt{2018ApJ...855...22G}). In a previous work, we simulated possible JWST programs to obtain the expected number of SNe that have multiple images behind the galaxy clusters A1689 and A370 (P16, P18). Here, we have considered five {\it Hubble Frontier Fields} galaxy clusters as gravitational telescopes, motivated by the recent publications of new improved cluster models (K18 and M18). We presented here the magnifications and time delays of the multiple images of the background galaxies based on the lensing models of K18 and M18. We simulated the light curves of SNe  that could explode in the multiply-imaged galaxies behind the galaxy clusters, and examined whether they will be detectable by the JWST/NIRCam bands F115W and F150W.  Considering F115W, we found that JWST will be sensitive to at least 75 pairs of SNe Ia with multiple images exploding in the multiply-imaged galaxies with time delays of less than 5 years (the nominal duration of JWST). For the most common type of CC~SNe, the SNe~IIP,  this number is slightly lower, 56, as they are on average fainter than SNe Ia. The same consideration, but for F150W gives better prospects, 93 (80) pairs of SNe~Ia (SNe~IIP) images because it is more appropriate for high-$z$ SNe.

We also simulated possible JWST surveys to obtain the expected number of lensed SNe that have multiple images, and concluded that A1689 is the lens that offers the best prospects. The high-$z$ SNe that will be discovered with the JWST can be spectroscopically followed with the 
 new generation facilities such as the Extremely Large Telescope \citep{2007Msngr.127...11G} and the Giant Magellan Telescope \citep{2012SPIE.8444E..1HJ} which will start operate in the 2020s. 

\bibliography{references}

\begin{thebibliography}{40}%
\makeatletter
\providecommand \@ifxundefined [1]{%
 \@ifx{#1\undefined}
}%
\providecommand \@ifnum [1]{%
 \ifnum #1\expandafter \@firstoftwo
 \else \expandafter \@secondoftwo
 \fi
}%
\providecommand \@ifx [1]{%
 \ifx #1\expandafter \@firstoftwo
 \else \expandafter \@secondoftwo
 \fi
}%
\providecommand \natexlab [1]{#1}%
\providecommand \enquote  [1]{``#1''}%
\providecommand \bibnamefont  [1]{#1}%
\providecommand \bibfnamefont [1]{#1}%
\providecommand \citenamefont [1]{#1}%
\providecommand \href@noop [0]{\@secondoftwo}%
\providecommand \href [0]{\begingroup \@sanitize@url \@href}%
\providecommand \@href[1]{\@@startlink{#1}\@@href}%
\providecommand \@@href[1]{\endgroup#1\@@endlink}%
\providecommand \@sanitize@url [0]{\catcode `\\12\catcode `\$12\catcode
  `\&12\catcode `\#12\catcode `\^12\catcode `\_12\catcode `\%12\relax}%
\providecommand \@@startlink[1]{}%
\providecommand \@@endlink[0]{}%
\providecommand \url  [0]{\begingroup\@sanitize@url \@url }%
\providecommand \@url [1]{\endgroup\@href {#1}{\urlprefix }}%
\providecommand \urlprefix  [0]{URL }%
\providecommand \Eprint [0]{\href }%
\providecommand \doibase [0]{http://dx.doi.org/}%
\providecommand \selectlanguage [0]{\@gobble}%
\providecommand \bibinfo  [0]{\@secondoftwo}%
\providecommand \bibfield  [0]{\@secondoftwo}%
\providecommand \translation [1]{[#1]}%
\providecommand \BibitemOpen [0]{}%
\providecommand \bibitemStop [0]{}%
\providecommand \bibitemNoStop [0]{.\EOS\space}%
\providecommand \EOS [0]{\spacefactor3000\relax}%
\providecommand \BibitemShut  [1]{\csname bibitem#1\endcsname}%
\let\auto@bib@innerbib\@empty
\bibitem [{\citenamefont {{Refsdal}}(1964)}]{1964MNRAS.128..307R}%
  \BibitemOpen
  \bibfield  {author} {\bibinfo {author} {\bibfnamefont {S.}~\bibnamefont
  {{Refsdal}}},\ }\href {\doibase 10.1093/mnras/128.4.307} {\bibfield
  {journal} {\bibinfo  {journal} {\mnras}\ }\textbf {\bibinfo {volume} {128}},\
  \bibinfo {pages} {307} (\bibinfo {year} {1964})}\BibitemShut {NoStop}%
\bibitem [{\citenamefont {{Treu}}\ and\ \citenamefont
  {{Marshall}}(2016)}]{2016A&ARv..24...11T}%
  \BibitemOpen
  \bibfield  {author} {\bibinfo {author} {\bibfnamefont {T.}~\bibnamefont
  {{Treu}}}\ and\ \bibinfo {author} {\bibfnamefont {P.~J.}\ \bibnamefont
  {{Marshall}}},\ }\href {\doibase 10.1007/s00159-016-0096-8} {\bibfield
  {journal} {\bibinfo  {journal} {\aapr}\ }\textbf {\bibinfo {volume} {24}},\
  \bibinfo {eid} {11} (\bibinfo {year} {2016})},\ \Eprint
  {http://arxiv.org/abs/1605.05333} {arXiv:1605.05333} \BibitemShut {NoStop}%
\bibitem [{\citenamefont {{Kelly}}\ \emph {et~al.}(2015)\citenamefont
  {{Kelly}}, \citenamefont {{Rodney}}, \citenamefont {{Treu}}, \citenamefont
  {{Foley}}, \citenamefont {{Brammer}}, \citenamefont {{Schmidt}},
  \citenamefont {{Zitrin}}, \citenamefont {{Sonnenfeld}}, \citenamefont
  {{Strolger}}, \citenamefont {{Graur}}, \citenamefont {{Filippenko}},
  \citenamefont {{Jha}}, \citenamefont {{Riess}}, \citenamefont {{Bradac}},
  \citenamefont {{Weiner}}, \citenamefont {{Scolnic}}, \citenamefont
  {{Malkan}}, \citenamefont {{von der Linden}}, \citenamefont {{Trenti}},
  \citenamefont {{Hjorth}}, \citenamefont {{Gavazzi}}, \citenamefont
  {{Fontana}}, \citenamefont {{Merten}}, \citenamefont {{McCully}},
  \citenamefont {{Jones}}, \citenamefont {{Postman}}, \citenamefont
  {{Dressler}}, \citenamefont {{Patel}}, \citenamefont {{Cenko}}, \citenamefont
  {{Graham}},\ and\ \citenamefont {{Tucker}}}]{2015Sci...347.1123K}%
  \BibitemOpen
  \bibfield  {author} {\bibinfo {author} {\bibfnamefont {P.~L.}\ \bibnamefont
  {{Kelly}}}, \bibinfo {author} {\bibfnamefont {S.~A.}\ \bibnamefont
  {{Rodney}}}, \bibinfo {author} {\bibfnamefont {T.}~\bibnamefont {{Treu}}},
  \bibinfo {author} {\bibfnamefont {R.~J.}\ \bibnamefont {{Foley}}}, \bibinfo
  {author} {\bibfnamefont {G.}~\bibnamefont {{Brammer}}}, \bibinfo {author}
  {\bibfnamefont {K.~B.}\ \bibnamefont {{Schmidt}}}, \bibinfo {author}
  {\bibfnamefont {A.}~\bibnamefont {{Zitrin}}}, \bibinfo {author}
  {\bibfnamefont {A.}~\bibnamefont {{Sonnenfeld}}}, \bibinfo {author}
  {\bibfnamefont {L.-G.}\ \bibnamefont {{Strolger}}}, \bibinfo {author}
  {\bibfnamefont {O.}~\bibnamefont {{Graur}}}, \bibinfo {author} {\bibfnamefont
  {A.~V.}\ \bibnamefont {{Filippenko}}}, \bibinfo {author} {\bibfnamefont
  {S.~W.}\ \bibnamefont {{Jha}}}, \bibinfo {author} {\bibfnamefont {A.~G.}\
  \bibnamefont {{Riess}}}, \bibinfo {author} {\bibfnamefont {M.}~\bibnamefont
  {{Bradac}}}, \bibinfo {author} {\bibfnamefont {B.~J.}\ \bibnamefont
  {{Weiner}}}, \bibinfo {author} {\bibfnamefont {D.}~\bibnamefont {{Scolnic}}},
  \bibinfo {author} {\bibfnamefont {M.~A.}\ \bibnamefont {{Malkan}}}, \bibinfo
  {author} {\bibfnamefont {A.}~\bibnamefont {{von der Linden}}}, \bibinfo
  {author} {\bibfnamefont {M.}~\bibnamefont {{Trenti}}}, \bibinfo {author}
  {\bibfnamefont {J.}~\bibnamefont {{Hjorth}}}, \bibinfo {author}
  {\bibfnamefont {R.}~\bibnamefont {{Gavazzi}}}, \bibinfo {author}
  {\bibfnamefont {A.}~\bibnamefont {{Fontana}}}, \bibinfo {author}
  {\bibfnamefont {J.~C.}\ \bibnamefont {{Merten}}}, \bibinfo {author}
  {\bibfnamefont {C.}~\bibnamefont {{McCully}}}, \bibinfo {author}
  {\bibfnamefont {T.}~\bibnamefont {{Jones}}}, \bibinfo {author} {\bibfnamefont
  {M.}~\bibnamefont {{Postman}}}, \bibinfo {author} {\bibfnamefont
  {A.}~\bibnamefont {{Dressler}}}, \bibinfo {author} {\bibfnamefont
  {B.}~\bibnamefont {{Patel}}}, \bibinfo {author} {\bibfnamefont {S.~B.}\
  \bibnamefont {{Cenko}}}, \bibinfo {author} {\bibfnamefont {M.~L.}\
  \bibnamefont {{Graham}}}, \ and\ \bibinfo {author} {\bibfnamefont {B.~E.}\
  \bibnamefont {{Tucker}}},\ }\href {\doibase 10.1126/science.aaa3350}
  {\bibfield  {journal} {\bibinfo  {journal} {Science}\ }\textbf {\bibinfo
  {volume} {347}},\ \bibinfo {pages} {1123} (\bibinfo {year} {2015})},\ \Eprint
  {http://arxiv.org/abs/1411.6009} {arXiv:1411.6009} \BibitemShut {NoStop}%
\bibitem [{\citenamefont {{Vega-Ferrero}}\ \emph {et~al.}(2018)\citenamefont
  {{Vega-Ferrero}}, \citenamefont {{Diego}}, \citenamefont {{Miranda}},\ and\
  \citenamefont {{Bernstein}}}]{2018ApJ...853L..31V}%
  \BibitemOpen
  \bibfield  {author} {\bibinfo {author} {\bibfnamefont {J.}~\bibnamefont
  {{Vega-Ferrero}}}, \bibinfo {author} {\bibfnamefont {J.~M.}\ \bibnamefont
  {{Diego}}}, \bibinfo {author} {\bibfnamefont {V.}~\bibnamefont {{Miranda}}},
  \ and\ \bibinfo {author} {\bibfnamefont {G.~M.}\ \bibnamefont
  {{Bernstein}}},\ }\href {\doibase 10.3847/2041-8213/aaa95f} {\bibfield
  {journal} {\bibinfo  {journal} {\apjl}\ }\textbf {\bibinfo {volume} {853}},\
  \bibinfo {eid} {L31} (\bibinfo {year} {2018})},\ \Eprint
  {http://arxiv.org/abs/1712.05800} {arXiv:1712.05800} \BibitemShut {NoStop}%
\bibitem [{\citenamefont {{Grillo}}\ \emph {et~al.}(2018)\citenamefont
  {{Grillo}}, \citenamefont {{Rosati}}, \citenamefont {{Suyu}}, \citenamefont
  {{Balestra}}, \citenamefont {{Caminha}}, \citenamefont {{Halkola}},
  \citenamefont {{Kelly}}, \citenamefont {{Lombardi}}, \citenamefont
  {{Mercurio}}, \citenamefont {{Rodney}},\ and\ \citenamefont
  {{Treu}}}]{2018ApJ...860...94G}%
  \BibitemOpen
  \bibfield  {author} {\bibinfo {author} {\bibfnamefont {C.}~\bibnamefont
  {{Grillo}}}, \bibinfo {author} {\bibfnamefont {P.}~\bibnamefont {{Rosati}}},
  \bibinfo {author} {\bibfnamefont {S.~H.}\ \bibnamefont {{Suyu}}}, \bibinfo
  {author} {\bibfnamefont {I.}~\bibnamefont {{Balestra}}}, \bibinfo {author}
  {\bibfnamefont {G.~B.}\ \bibnamefont {{Caminha}}}, \bibinfo {author}
  {\bibfnamefont {A.}~\bibnamefont {{Halkola}}}, \bibinfo {author}
  {\bibfnamefont {P.~L.}\ \bibnamefont {{Kelly}}}, \bibinfo {author}
  {\bibfnamefont {M.}~\bibnamefont {{Lombardi}}}, \bibinfo {author}
  {\bibfnamefont {A.}~\bibnamefont {{Mercurio}}}, \bibinfo {author}
  {\bibfnamefont {S.~A.}\ \bibnamefont {{Rodney}}}, \ and\ \bibinfo {author}
  {\bibfnamefont {T.}~\bibnamefont {{Treu}}},\ }\href {\doibase
  10.3847/1538-4357/aac2c9} {\bibfield  {journal} {\bibinfo  {journal} {\apj}\
  }\textbf {\bibinfo {volume} {860}},\ \bibinfo {eid} {94} (\bibinfo {year}
  {2018})},\ \Eprint {http://arxiv.org/abs/1802.01584} {arXiv:1802.01584}
  \BibitemShut {NoStop}%
\bibitem [{\citenamefont {{Planck Collaboration}}\ \emph
  {et~al.}(2016)\citenamefont {{Planck Collaboration}}, \citenamefont {{Ade}},
  \citenamefont {{Aghanim}}, \citenamefont {{Arnaud}}, \citenamefont
  {{Ashdown}}, \citenamefont {{Aumont}}, \citenamefont {{Baccigalupi}},
  \citenamefont {{Banday}}, \citenamefont {{Barreiro}}, \citenamefont
  {{Bartlett}},\ and\ \citenamefont {et~al.}}]{2016A&A...594A..13P}%
  \BibitemOpen
  \bibfield  {author} {\bibinfo {author} {\bibnamefont {{Planck
  Collaboration}}}, \bibinfo {author} {\bibfnamefont {P.~A.~R.}\ \bibnamefont
  {{Ade}}}, \bibinfo {author} {\bibfnamefont {N.}~\bibnamefont {{Aghanim}}},
  \bibinfo {author} {\bibfnamefont {M.}~\bibnamefont {{Arnaud}}}, \bibinfo
  {author} {\bibfnamefont {M.}~\bibnamefont {{Ashdown}}}, \bibinfo {author}
  {\bibfnamefont {J.}~\bibnamefont {{Aumont}}}, \bibinfo {author}
  {\bibfnamefont {C.}~\bibnamefont {{Baccigalupi}}}, \bibinfo {author}
  {\bibfnamefont {A.~J.}\ \bibnamefont {{Banday}}}, \bibinfo {author}
  {\bibfnamefont {R.~B.}\ \bibnamefont {{Barreiro}}}, \bibinfo {author}
  {\bibfnamefont {J.~G.}\ \bibnamefont {{Bartlett}}}, \ and\ \bibinfo {author}
  {\bibnamefont {et~al.}},\ }\href {\doibase 10.1051/0004-6361/201525830}
  {\bibfield  {journal} {\bibinfo  {journal} {\aap}\ }\textbf {\bibinfo
  {volume} {594}},\ \bibinfo {eid} {A13} (\bibinfo {year} {2016})},\ \Eprint
  {http://arxiv.org/abs/1502.01589} {arXiv:1502.01589} \BibitemShut {NoStop}%
\bibitem [{\citenamefont {{Riess}}\ \emph {et~al.}(2016)\citenamefont
  {{Riess}}, \citenamefont {{Macri}}, \citenamefont {{Hoffmann}}, \citenamefont
  {{Scolnic}}, \citenamefont {{Casertano}}, \citenamefont {{Filippenko}},
  \citenamefont {{Tucker}}, \citenamefont {{Reid}}, \citenamefont {{Jones}},
  \citenamefont {{Silverman}}, \citenamefont {{Chornock}}, \citenamefont
  {{Challis}}, \citenamefont {{Yuan}}, \citenamefont {{Brown}},\ and\
  \citenamefont {{Foley}}}]{2016ApJ...826...56R}%
  \BibitemOpen
  \bibfield  {author} {\bibinfo {author} {\bibfnamefont {A.~G.}\ \bibnamefont
  {{Riess}}}, \bibinfo {author} {\bibfnamefont {L.~M.}\ \bibnamefont
  {{Macri}}}, \bibinfo {author} {\bibfnamefont {S.~L.}\ \bibnamefont
  {{Hoffmann}}}, \bibinfo {author} {\bibfnamefont {D.}~\bibnamefont
  {{Scolnic}}}, \bibinfo {author} {\bibfnamefont {S.}~\bibnamefont
  {{Casertano}}}, \bibinfo {author} {\bibfnamefont {A.~V.}\ \bibnamefont
  {{Filippenko}}}, \bibinfo {author} {\bibfnamefont {B.~E.}\ \bibnamefont
  {{Tucker}}}, \bibinfo {author} {\bibfnamefont {M.~J.}\ \bibnamefont
  {{Reid}}}, \bibinfo {author} {\bibfnamefont {D.~O.}\ \bibnamefont {{Jones}}},
  \bibinfo {author} {\bibfnamefont {J.~M.}\ \bibnamefont {{Silverman}}},
  \bibinfo {author} {\bibfnamefont {R.}~\bibnamefont {{Chornock}}}, \bibinfo
  {author} {\bibfnamefont {P.}~\bibnamefont {{Challis}}}, \bibinfo {author}
  {\bibfnamefont {W.}~\bibnamefont {{Yuan}}}, \bibinfo {author} {\bibfnamefont
  {P.~J.}\ \bibnamefont {{Brown}}}, \ and\ \bibinfo {author} {\bibfnamefont
  {R.~J.}\ \bibnamefont {{Foley}}},\ }\href {\doibase
  10.3847/0004-637X/826/1/56} {\bibfield  {journal} {\bibinfo  {journal}
  {\apj}\ }\textbf {\bibinfo {volume} {826}},\ \bibinfo {eid} {56} (\bibinfo
  {year} {2016})},\ \Eprint {http://arxiv.org/abs/1604.01424}
  {arXiv:1604.01424} \BibitemShut {NoStop}%
\bibitem [{\citenamefont {{Riess}}\ \emph {et~al.}(2018)\citenamefont
  {{Riess}}, \citenamefont {{Casertano}}, \citenamefont {{Yuan}}, \citenamefont
  {{Macri}}, \citenamefont {{Anderson}}, \citenamefont {{MacKenty}},
  \citenamefont {{Bowers}}, \citenamefont {{Clubb}}, \citenamefont
  {{Filippenko}}, \citenamefont {{Jones}},\ and\ \citenamefont
  {{Tucker}}}]{2018ApJ...855..136R}%
  \BibitemOpen
  \bibfield  {author} {\bibinfo {author} {\bibfnamefont {A.~G.}\ \bibnamefont
  {{Riess}}}, \bibinfo {author} {\bibfnamefont {S.}~\bibnamefont
  {{Casertano}}}, \bibinfo {author} {\bibfnamefont {W.}~\bibnamefont {{Yuan}}},
  \bibinfo {author} {\bibfnamefont {L.}~\bibnamefont {{Macri}}}, \bibinfo
  {author} {\bibfnamefont {J.}~\bibnamefont {{Anderson}}}, \bibinfo {author}
  {\bibfnamefont {J.~W.}\ \bibnamefont {{MacKenty}}}, \bibinfo {author}
  {\bibfnamefont {J.~B.}\ \bibnamefont {{Bowers}}}, \bibinfo {author}
  {\bibfnamefont {K.~I.}\ \bibnamefont {{Clubb}}}, \bibinfo {author}
  {\bibfnamefont {A.~V.}\ \bibnamefont {{Filippenko}}}, \bibinfo {author}
  {\bibfnamefont {D.~O.}\ \bibnamefont {{Jones}}}, \ and\ \bibinfo {author}
  {\bibfnamefont {B.~E.}\ \bibnamefont {{Tucker}}},\ }\href {\doibase
  10.3847/1538-4357/aaadb7} {\bibfield  {journal} {\bibinfo  {journal} {\apj}\
  }\textbf {\bibinfo {volume} {855}},\ \bibinfo {eid} {136} (\bibinfo {year}
  {2018})},\ \Eprint {http://arxiv.org/abs/1801.01120} {arXiv:1801.01120
  [astro-ph.SR]} \BibitemShut {NoStop}%
\bibitem [{\citenamefont {{Gunnarsson}}\ and\ \citenamefont
  {{Goobar}}(2003)}]{2003A&A...405..859G}%
  \BibitemOpen
  \bibfield  {author} {\bibinfo {author} {\bibfnamefont {C.}~\bibnamefont
  {{Gunnarsson}}}\ and\ \bibinfo {author} {\bibfnamefont {A.}~\bibnamefont
  {{Goobar}}},\ }\href {\doibase 10.1051/0004-6361:20030648} {\bibfield
  {journal} {\bibinfo  {journal} {\aap}\ }\textbf {\bibinfo {volume} {405}},\
  \bibinfo {pages} {859} (\bibinfo {year} {2003})},\ \Eprint
  {http://arxiv.org/abs/astro-ph/0211401} {astro-ph/0211401} \BibitemShut
  {NoStop}%
\bibitem [{\citenamefont {{Suyu}}\ \emph {et~al.}(2017)\citenamefont {{Suyu}},
  \citenamefont {{Bonvin}}, \citenamefont {{Courbin}}, \citenamefont
  {{Fassnacht}}, \citenamefont {{Rusu}}, \citenamefont {{Sluse}}, \citenamefont
  {{Treu}}, \citenamefont {{Wong}}, \citenamefont {{Auger}}, \citenamefont
  {{Ding}}, \citenamefont {{Hilbert}}, \citenamefont {{Marshall}},
  \citenamefont {{Rumbaugh}}, \citenamefont {{Sonnenfeld}}, \citenamefont
  {{Tewes}}, \citenamefont {{Tihhonova}}, \citenamefont {{Agnello}},
  \citenamefont {{Blandford}}, \citenamefont {{Chen}}, \citenamefont
  {{Collett}}, \citenamefont {{Koopmans}}, \citenamefont {{Liao}},
  \citenamefont {{Meylan}},\ and\ \citenamefont
  {{Spiniello}}}]{2017MNRAS.468.2590S}%
  \BibitemOpen
  \bibfield  {author} {\bibinfo {author} {\bibfnamefont {S.~H.}\ \bibnamefont
  {{Suyu}}}, \bibinfo {author} {\bibfnamefont {V.}~\bibnamefont {{Bonvin}}},
  \bibinfo {author} {\bibfnamefont {F.}~\bibnamefont {{Courbin}}}, \bibinfo
  {author} {\bibfnamefont {C.~D.}\ \bibnamefont {{Fassnacht}}}, \bibinfo
  {author} {\bibfnamefont {C.~E.}\ \bibnamefont {{Rusu}}}, \bibinfo {author}
  {\bibfnamefont {D.}~\bibnamefont {{Sluse}}}, \bibinfo {author} {\bibfnamefont
  {T.}~\bibnamefont {{Treu}}}, \bibinfo {author} {\bibfnamefont {K.~C.}\
  \bibnamefont {{Wong}}}, \bibinfo {author} {\bibfnamefont {M.~W.}\
  \bibnamefont {{Auger}}}, \bibinfo {author} {\bibfnamefont {X.}~\bibnamefont
  {{Ding}}}, \bibinfo {author} {\bibfnamefont {S.}~\bibnamefont {{Hilbert}}},
  \bibinfo {author} {\bibfnamefont {P.~J.}\ \bibnamefont {{Marshall}}},
  \bibinfo {author} {\bibfnamefont {N.}~\bibnamefont {{Rumbaugh}}}, \bibinfo
  {author} {\bibfnamefont {A.}~\bibnamefont {{Sonnenfeld}}}, \bibinfo {author}
  {\bibfnamefont {M.}~\bibnamefont {{Tewes}}}, \bibinfo {author} {\bibfnamefont
  {O.}~\bibnamefont {{Tihhonova}}}, \bibinfo {author} {\bibfnamefont
  {A.}~\bibnamefont {{Agnello}}}, \bibinfo {author} {\bibfnamefont {R.~D.}\
  \bibnamefont {{Blandford}}}, \bibinfo {author} {\bibfnamefont {G.~C.-F.}\
  \bibnamefont {{Chen}}}, \bibinfo {author} {\bibfnamefont {T.}~\bibnamefont
  {{Collett}}}, \bibinfo {author} {\bibfnamefont {L.~V.~E.}\ \bibnamefont
  {{Koopmans}}}, \bibinfo {author} {\bibfnamefont {K.}~\bibnamefont {{Liao}}},
  \bibinfo {author} {\bibfnamefont {G.}~\bibnamefont {{Meylan}}}, \ and\
  \bibinfo {author} {\bibfnamefont {C.}~\bibnamefont {{Spiniello}}},\ }\href
  {\doibase 10.1093/mnras/stx483} {\bibfield  {journal} {\bibinfo  {journal}
  {\mnras}\ }\textbf {\bibinfo {volume} {468}},\ \bibinfo {pages} {2590}
  (\bibinfo {year} {2017})},\ \Eprint {http://arxiv.org/abs/1607.00017}
  {arXiv:1607.00017} \BibitemShut {NoStop}%
\bibitem [{\citenamefont {{Oguri}}\ and\ \citenamefont
  {{Kawano}}(2003)}]{2003MNRAS.338L..25O}%
  \BibitemOpen
  \bibfield  {author} {\bibinfo {author} {\bibfnamefont {M.}~\bibnamefont
  {{Oguri}}}\ and\ \bibinfo {author} {\bibfnamefont {Y.}~\bibnamefont
  {{Kawano}}},\ }\href {\doibase 10.1046/j.1365-8711.2003.06290.x} {\bibfield
  {journal} {\bibinfo  {journal} {\mnras}\ }\textbf {\bibinfo {volume} {338}},\
  \bibinfo {pages} {L25} (\bibinfo {year} {2003})},\ \Eprint
  {http://arxiv.org/abs/astro-ph/0211499} {astro-ph/0211499} \BibitemShut
  {NoStop}%
\bibitem [{\citenamefont {{Patel}}\ \emph {et~al.}(2014)\citenamefont
  {{Patel}}, \citenamefont {{McCully}}, \citenamefont {{Jha}}, \citenamefont
  {{Rodney}}, \citenamefont {{Jones}}, \citenamefont {{Graur}}, \citenamefont
  {{Merten}}, \citenamefont {{Zitrin}}, \citenamefont {{Riess}}, \citenamefont
  {{Matheson}}, \citenamefont {{Sako}}, \citenamefont {{Holoien}},
  \citenamefont {{Postman}}, \citenamefont {{Coe}}, \citenamefont
  {{Bartelmann}}, \citenamefont {{Balestra}}, \citenamefont {{Ben{\'{\i}}tez}},
  \citenamefont {{Bouwens}}, \citenamefont {{Bradley}}, \citenamefont
  {{Broadhurst}}, \citenamefont {{Cenko}}, \citenamefont {{Donahue}},
  \citenamefont {{Filippenko}}, \citenamefont {{Ford}}, \citenamefont
  {{Garnavich}}, \citenamefont {{Grillo}}, \citenamefont {{Infante}},
  \citenamefont {{Jouvel}}, \citenamefont {{Kelson}}, \citenamefont
  {{Koekemoer}}, \citenamefont {{Lahav}}, \citenamefont {{Lemze}},
  \citenamefont {{Maoz}}, \citenamefont {{Medezinski}}, \citenamefont
  {{Melchior}}, \citenamefont {{Meneghetti}}, \citenamefont {{Molino}},
  \citenamefont {{Moustakas}}, \citenamefont {{Moustakas}}, \citenamefont
  {{Nonino}}, \citenamefont {{Rosati}}, \citenamefont {{Seitz}}, \citenamefont
  {{Strolger}}, \citenamefont {{Umetsu}},\ and\ \citenamefont
  {{Zheng}}}]{2014ApJ...786....9P}%
  \BibitemOpen
  \bibfield  {author} {\bibinfo {author} {\bibfnamefont {B.}~\bibnamefont
  {{Patel}}}, \bibinfo {author} {\bibfnamefont {C.}~\bibnamefont {{McCully}}},
  \bibinfo {author} {\bibfnamefont {S.~W.}\ \bibnamefont {{Jha}}}, \bibinfo
  {author} {\bibfnamefont {S.~A.}\ \bibnamefont {{Rodney}}}, \bibinfo {author}
  {\bibfnamefont {D.~O.}\ \bibnamefont {{Jones}}}, \bibinfo {author}
  {\bibfnamefont {O.}~\bibnamefont {{Graur}}}, \bibinfo {author} {\bibfnamefont
  {J.}~\bibnamefont {{Merten}}}, \bibinfo {author} {\bibfnamefont
  {A.}~\bibnamefont {{Zitrin}}}, \bibinfo {author} {\bibfnamefont {A.~G.}\
  \bibnamefont {{Riess}}}, \bibinfo {author} {\bibfnamefont {T.}~\bibnamefont
  {{Matheson}}}, \bibinfo {author} {\bibfnamefont {M.}~\bibnamefont {{Sako}}},
  \bibinfo {author} {\bibfnamefont {T.~W.-S.}\ \bibnamefont {{Holoien}}},
  \bibinfo {author} {\bibfnamefont {M.}~\bibnamefont {{Postman}}}, \bibinfo
  {author} {\bibfnamefont {D.}~\bibnamefont {{Coe}}}, \bibinfo {author}
  {\bibfnamefont {M.}~\bibnamefont {{Bartelmann}}}, \bibinfo {author}
  {\bibfnamefont {I.}~\bibnamefont {{Balestra}}}, \bibinfo {author}
  {\bibfnamefont {N.}~\bibnamefont {{Ben{\'{\i}}tez}}}, \bibinfo {author}
  {\bibfnamefont {R.}~\bibnamefont {{Bouwens}}}, \bibinfo {author}
  {\bibfnamefont {L.}~\bibnamefont {{Bradley}}}, \bibinfo {author}
  {\bibfnamefont {T.}~\bibnamefont {{Broadhurst}}}, \bibinfo {author}
  {\bibfnamefont {S.~B.}\ \bibnamefont {{Cenko}}}, \bibinfo {author}
  {\bibfnamefont {M.}~\bibnamefont {{Donahue}}}, \bibinfo {author}
  {\bibfnamefont {A.~V.}\ \bibnamefont {{Filippenko}}}, \bibinfo {author}
  {\bibfnamefont {H.}~\bibnamefont {{Ford}}}, \bibinfo {author} {\bibfnamefont
  {P.}~\bibnamefont {{Garnavich}}}, \bibinfo {author} {\bibfnamefont
  {C.}~\bibnamefont {{Grillo}}}, \bibinfo {author} {\bibfnamefont
  {L.}~\bibnamefont {{Infante}}}, \bibinfo {author} {\bibfnamefont
  {S.}~\bibnamefont {{Jouvel}}}, \bibinfo {author} {\bibfnamefont
  {D.}~\bibnamefont {{Kelson}}}, \bibinfo {author} {\bibfnamefont
  {A.}~\bibnamefont {{Koekemoer}}}, \bibinfo {author} {\bibfnamefont
  {O.}~\bibnamefont {{Lahav}}}, \bibinfo {author} {\bibfnamefont
  {D.}~\bibnamefont {{Lemze}}}, \bibinfo {author} {\bibfnamefont
  {D.}~\bibnamefont {{Maoz}}}, \bibinfo {author} {\bibfnamefont
  {E.}~\bibnamefont {{Medezinski}}}, \bibinfo {author} {\bibfnamefont
  {P.}~\bibnamefont {{Melchior}}}, \bibinfo {author} {\bibfnamefont
  {M.}~\bibnamefont {{Meneghetti}}}, \bibinfo {author} {\bibfnamefont
  {A.}~\bibnamefont {{Molino}}}, \bibinfo {author} {\bibfnamefont
  {J.}~\bibnamefont {{Moustakas}}}, \bibinfo {author} {\bibfnamefont {L.~A.}\
  \bibnamefont {{Moustakas}}}, \bibinfo {author} {\bibfnamefont
  {M.}~\bibnamefont {{Nonino}}}, \bibinfo {author} {\bibfnamefont
  {P.}~\bibnamefont {{Rosati}}}, \bibinfo {author} {\bibfnamefont
  {S.}~\bibnamefont {{Seitz}}}, \bibinfo {author} {\bibfnamefont {L.~G.}\
  \bibnamefont {{Strolger}}}, \bibinfo {author} {\bibfnamefont
  {K.}~\bibnamefont {{Umetsu}}}, \ and\ \bibinfo {author} {\bibfnamefont
  {W.}~\bibnamefont {{Zheng}}},\ }\href {\doibase 10.1088/0004-637X/786/1/9}
  {\bibfield  {journal} {\bibinfo  {journal} {\apj}\ }\textbf {\bibinfo
  {volume} {786}},\ \bibinfo {eid} {9} (\bibinfo {year} {2014})},\ \Eprint
  {http://arxiv.org/abs/1312.0943} {arXiv:1312.0943 [astro-ph.CO]} \BibitemShut
  {NoStop}%
\bibitem [{\citenamefont {{Nordin}}\ \emph {et~al.}(2014)\citenamefont
  {{Nordin}}, \citenamefont {{Rubin}}, \citenamefont {{Richard}}, \citenamefont
  {{Rykoff}}, \citenamefont {{Aldering}}, \citenamefont {{Amanullah}},
  \citenamefont {{Atek}}, \citenamefont {{Barbary}}, \citenamefont {{Deustua}},
  \citenamefont {{Fakhouri}}, \citenamefont {{Fruchter}}, \citenamefont
  {{Goobar}}, \citenamefont {{Hook}}, \citenamefont {{Hsiao}}, \citenamefont
  {{Huang}}, \citenamefont {{Kneib}}, \citenamefont {{Lidman}}, \citenamefont
  {{Meyers}}, \citenamefont {{Perlmutter}}, \citenamefont {{Saunders}},
  \citenamefont {{Spadafora}}, \citenamefont {{Suzuki}},\ and\ \citenamefont
  {{Supernova Cosmology Project}}}]{2014MNRAS.440.2742N}%
  \BibitemOpen
  \bibfield  {author} {\bibinfo {author} {\bibfnamefont {J.}~\bibnamefont
  {{Nordin}}}, \bibinfo {author} {\bibfnamefont {D.}~\bibnamefont {{Rubin}}},
  \bibinfo {author} {\bibfnamefont {J.}~\bibnamefont {{Richard}}}, \bibinfo
  {author} {\bibfnamefont {E.}~\bibnamefont {{Rykoff}}}, \bibinfo {author}
  {\bibfnamefont {G.}~\bibnamefont {{Aldering}}}, \bibinfo {author}
  {\bibfnamefont {R.}~\bibnamefont {{Amanullah}}}, \bibinfo {author}
  {\bibfnamefont {H.}~\bibnamefont {{Atek}}}, \bibinfo {author} {\bibfnamefont
  {K.}~\bibnamefont {{Barbary}}}, \bibinfo {author} {\bibfnamefont
  {S.}~\bibnamefont {{Deustua}}}, \bibinfo {author} {\bibfnamefont {H.~K.}\
  \bibnamefont {{Fakhouri}}}, \bibinfo {author} {\bibfnamefont {A.~S.}\
  \bibnamefont {{Fruchter}}}, \bibinfo {author} {\bibfnamefont
  {A.}~\bibnamefont {{Goobar}}}, \bibinfo {author} {\bibfnamefont
  {I.}~\bibnamefont {{Hook}}}, \bibinfo {author} {\bibfnamefont {E.~Y.}\
  \bibnamefont {{Hsiao}}}, \bibinfo {author} {\bibfnamefont {X.}~\bibnamefont
  {{Huang}}}, \bibinfo {author} {\bibfnamefont {J.-P.}\ \bibnamefont
  {{Kneib}}}, \bibinfo {author} {\bibfnamefont {C.}~\bibnamefont {{Lidman}}},
  \bibinfo {author} {\bibfnamefont {J.}~\bibnamefont {{Meyers}}}, \bibinfo
  {author} {\bibfnamefont {S.}~\bibnamefont {{Perlmutter}}}, \bibinfo {author}
  {\bibfnamefont {C.}~\bibnamefont {{Saunders}}}, \bibinfo {author}
  {\bibfnamefont {A.~L.}\ \bibnamefont {{Spadafora}}}, \bibinfo {author}
  {\bibfnamefont {N.}~\bibnamefont {{Suzuki}}}, \ and\ \bibinfo {author}
  {\bibnamefont {{Supernova Cosmology Project}}},\ }\href {\doibase
  10.1093/mnras/stu376} {\bibfield  {journal} {\bibinfo  {journal} {\mnras}\
  }\textbf {\bibinfo {volume} {440}},\ \bibinfo {pages} {2742} (\bibinfo {year}
  {2014})},\ \Eprint {http://arxiv.org/abs/1312.2576} {arXiv:1312.2576}
  \BibitemShut {NoStop}%
\bibitem [{\citenamefont {{Rodney}}\ \emph {et~al.}(2015)\citenamefont
  {{Rodney}}, \citenamefont {{Patel}}, \citenamefont {{Scolnic}}, \citenamefont
  {{Foley}}, \citenamefont {{Molino}}, \citenamefont {{Brammer}}, \citenamefont
  {{Jauzac}}, \citenamefont {{Brada{\v c}}}, \citenamefont {{Broadhurst}},
  \citenamefont {{Coe}}, \citenamefont {{Diego}}, \citenamefont {{Graur}},
  \citenamefont {{Hjorth}}, \citenamefont {{Hoag}}, \citenamefont {{Jha}},
  \citenamefont {{Johnson}}, \citenamefont {{Kelly}}, \citenamefont {{Lam}},
  \citenamefont {{McCully}}, \citenamefont {{Medezinski}}, \citenamefont
  {{Meneghetti}}, \citenamefont {{Merten}}, \citenamefont {{Richard}},
  \citenamefont {{Riess}}, \citenamefont {{Sharon}}, \citenamefont
  {{Strolger}}, \citenamefont {{Treu}}, \citenamefont {{Wang}}, \citenamefont
  {{Williams}},\ and\ \citenamefont {{Zitrin}}}]{2015ApJ...811...70R}%
  \BibitemOpen
  \bibfield  {author} {\bibinfo {author} {\bibfnamefont {S.~A.}\ \bibnamefont
  {{Rodney}}}, \bibinfo {author} {\bibfnamefont {B.}~\bibnamefont {{Patel}}},
  \bibinfo {author} {\bibfnamefont {D.}~\bibnamefont {{Scolnic}}}, \bibinfo
  {author} {\bibfnamefont {R.~J.}\ \bibnamefont {{Foley}}}, \bibinfo {author}
  {\bibfnamefont {A.}~\bibnamefont {{Molino}}}, \bibinfo {author}
  {\bibfnamefont {G.}~\bibnamefont {{Brammer}}}, \bibinfo {author}
  {\bibfnamefont {M.}~\bibnamefont {{Jauzac}}}, \bibinfo {author}
  {\bibfnamefont {M.}~\bibnamefont {{Brada{\v c}}}}, \bibinfo {author}
  {\bibfnamefont {T.}~\bibnamefont {{Broadhurst}}}, \bibinfo {author}
  {\bibfnamefont {D.}~\bibnamefont {{Coe}}}, \bibinfo {author} {\bibfnamefont
  {J.~M.}\ \bibnamefont {{Diego}}}, \bibinfo {author} {\bibfnamefont
  {O.}~\bibnamefont {{Graur}}}, \bibinfo {author} {\bibfnamefont
  {J.}~\bibnamefont {{Hjorth}}}, \bibinfo {author} {\bibfnamefont
  {A.}~\bibnamefont {{Hoag}}}, \bibinfo {author} {\bibfnamefont {S.~W.}\
  \bibnamefont {{Jha}}}, \bibinfo {author} {\bibfnamefont {T.~L.}\ \bibnamefont
  {{Johnson}}}, \bibinfo {author} {\bibfnamefont {P.}~\bibnamefont {{Kelly}}},
  \bibinfo {author} {\bibfnamefont {D.}~\bibnamefont {{Lam}}}, \bibinfo
  {author} {\bibfnamefont {C.}~\bibnamefont {{McCully}}}, \bibinfo {author}
  {\bibfnamefont {E.}~\bibnamefont {{Medezinski}}}, \bibinfo {author}
  {\bibfnamefont {M.}~\bibnamefont {{Meneghetti}}}, \bibinfo {author}
  {\bibfnamefont {J.}~\bibnamefont {{Merten}}}, \bibinfo {author}
  {\bibfnamefont {J.}~\bibnamefont {{Richard}}}, \bibinfo {author}
  {\bibfnamefont {A.}~\bibnamefont {{Riess}}}, \bibinfo {author} {\bibfnamefont
  {K.}~\bibnamefont {{Sharon}}}, \bibinfo {author} {\bibfnamefont {L.-G.}\
  \bibnamefont {{Strolger}}}, \bibinfo {author} {\bibfnamefont
  {T.}~\bibnamefont {{Treu}}}, \bibinfo {author} {\bibfnamefont
  {X.}~\bibnamefont {{Wang}}}, \bibinfo {author} {\bibfnamefont {L.~L.~R.}\
  \bibnamefont {{Williams}}}, \ and\ \bibinfo {author} {\bibfnamefont
  {A.}~\bibnamefont {{Zitrin}}},\ }\href {\doibase 10.1088/0004-637X/811/1/70}
  {\bibfield  {journal} {\bibinfo  {journal} {\apj}\ }\textbf {\bibinfo
  {volume} {811}},\ \bibinfo {eid} {70} (\bibinfo {year} {2015})},\ \Eprint
  {http://arxiv.org/abs/1505.06211} {arXiv:1505.06211} \BibitemShut {NoStop}%
\bibitem [{\citenamefont {{Petrushevska}}\ \emph {et~al.}(2017)\citenamefont
  {{Petrushevska}}, \citenamefont {{Amanullah}}, \citenamefont {{Bulla}},
  \citenamefont {{Kromer}}, \citenamefont {{Ferretti}}, \citenamefont
  {{Goobar}},\ and\ \citenamefont {{Papadogiannakis}}}]{2017A&A...603A.136P}%
  \BibitemOpen
  \bibfield  {author} {\bibinfo {author} {\bibfnamefont {T.}~\bibnamefont
  {{Petrushevska}}}, \bibinfo {author} {\bibfnamefont {R.}~\bibnamefont
  {{Amanullah}}}, \bibinfo {author} {\bibfnamefont {M.}~\bibnamefont
  {{Bulla}}}, \bibinfo {author} {\bibfnamefont {M.}~\bibnamefont {{Kromer}}},
  \bibinfo {author} {\bibfnamefont {R.}~\bibnamefont {{Ferretti}}}, \bibinfo
  {author} {\bibfnamefont {A.}~\bibnamefont {{Goobar}}}, \ and\ \bibinfo
  {author} {\bibfnamefont {S.}~\bibnamefont {{Papadogiannakis}}},\ }\href
  {\doibase 10.1051/0004-6361/201730989} {\bibfield  {journal} {\bibinfo
  {journal} {\aap}\ }\textbf {\bibinfo {volume} {603}},\ \bibinfo {eid} {A136}
  (\bibinfo {year} {2017})},\ \Eprint {http://arxiv.org/abs/1706.03770}
  {arXiv:1706.03770 [astro-ph.HE]} \BibitemShut {NoStop}%
\bibitem [{\citenamefont {{Sullivan}}\ \emph {et~al.}(2000)\citenamefont
  {{Sullivan}}, \citenamefont {{Ellis}}, \citenamefont {{Nugent}},
  \citenamefont {{Smail}},\ and\ \citenamefont
  {{Madau}}}]{2000MNRAS.319..549S}%
  \BibitemOpen
  \bibfield  {author} {\bibinfo {author} {\bibfnamefont {M.}~\bibnamefont
  {{Sullivan}}}, \bibinfo {author} {\bibfnamefont {R.}~\bibnamefont {{Ellis}}},
  \bibinfo {author} {\bibfnamefont {P.}~\bibnamefont {{Nugent}}}, \bibinfo
  {author} {\bibfnamefont {I.}~\bibnamefont {{Smail}}}, \ and\ \bibinfo
  {author} {\bibfnamefont {P.}~\bibnamefont {{Madau}}},\ }\href {\doibase
  10.1046/j.1365-8711.2000.03875.x} {\bibfield  {journal} {\bibinfo  {journal}
  {\mnras}\ }\textbf {\bibinfo {volume} {319}},\ \bibinfo {pages} {549}
  (\bibinfo {year} {2000})},\ \Eprint {http://arxiv.org/abs/astro-ph/0007228}
  {astro-ph/0007228} \BibitemShut {NoStop}%
\bibitem [{\citenamefont {{Petrushevska}}\ \emph {et~al.}(2016)\citenamefont
  {{Petrushevska}}, \citenamefont {{Amanullah}}, \citenamefont {{Goobar}},
  \citenamefont {{Fabbro}}, \citenamefont {{Johansson}}, \citenamefont
  {{Kjellsson}}, \citenamefont {{Lidman}}, \citenamefont {{Paech}},
  \citenamefont {{Richard}}, \citenamefont {{Dahle}}, \citenamefont
  {{Ferretti}}, \citenamefont {{Kneib}}, \citenamefont {{Limousin}},
  \citenamefont {{Nordin}},\ and\ \citenamefont
  {{Stanishev}}}]{2016A&A...594A..54P}%
  \BibitemOpen
  \bibfield  {author} {\bibinfo {author} {\bibfnamefont {T.}~\bibnamefont
  {{Petrushevska}}}, \bibinfo {author} {\bibfnamefont {R.}~\bibnamefont
  {{Amanullah}}}, \bibinfo {author} {\bibfnamefont {A.}~\bibnamefont
  {{Goobar}}}, \bibinfo {author} {\bibfnamefont {S.}~\bibnamefont {{Fabbro}}},
  \bibinfo {author} {\bibfnamefont {J.}~\bibnamefont {{Johansson}}}, \bibinfo
  {author} {\bibfnamefont {T.}~\bibnamefont {{Kjellsson}}}, \bibinfo {author}
  {\bibfnamefont {C.}~\bibnamefont {{Lidman}}}, \bibinfo {author}
  {\bibfnamefont {K.}~\bibnamefont {{Paech}}}, \bibinfo {author} {\bibfnamefont
  {J.}~\bibnamefont {{Richard}}}, \bibinfo {author} {\bibfnamefont
  {H.}~\bibnamefont {{Dahle}}}, \bibinfo {author} {\bibfnamefont
  {R.}~\bibnamefont {{Ferretti}}}, \bibinfo {author} {\bibfnamefont {J.~P.}\
  \bibnamefont {{Kneib}}}, \bibinfo {author} {\bibfnamefont {M.}~\bibnamefont
  {{Limousin}}}, \bibinfo {author} {\bibfnamefont {J.}~\bibnamefont
  {{Nordin}}}, \ and\ \bibinfo {author} {\bibfnamefont {V.}~\bibnamefont
  {{Stanishev}}},\ }\href {\doibase 10.1051/0004-6361/201628925} {\bibfield
  {journal} {\bibinfo  {journal} {\aap}\ }\textbf {\bibinfo {volume} {594}},\
  \bibinfo {eid} {A54} (\bibinfo {year} {2016})},\ \Eprint
  {http://arxiv.org/abs/1607.01617} {arXiv:1607.01617} \BibitemShut {NoStop}%
\bibitem [{\citenamefont {{Lotz}}\ \emph {et~al.}(2017)\citenamefont {{Lotz}},
  \citenamefont {{Koekemoer}}, \citenamefont {{Coe}}, \citenamefont {{Grogin}},
  \citenamefont {{Capak}}, \citenamefont {{Mack}}, \citenamefont {{Anderson}},
  \citenamefont {{Avila}}, \citenamefont {{Barker}}, \citenamefont
  {{Borncamp}}, \citenamefont {{Brammer}}, \citenamefont {{Durbin}},
  \citenamefont {{Gunning}}, \citenamefont {{Hilbert}}, \citenamefont
  {{Jenkner}}, \citenamefont {{Khandrika}}, \citenamefont {{Levay}},
  \citenamefont {{Lucas}}, \citenamefont {{MacKenty}}, \citenamefont {{Ogaz}},
  \citenamefont {{Porterfield}}, \citenamefont {{Reid}}, \citenamefont
  {{Robberto}}, \citenamefont {{Royle}}, \citenamefont {{Smith}}, \citenamefont
  {{Storrie-Lombardi}}, \citenamefont {{Sunnquist}}, \citenamefont {{Surace}},
  \citenamefont {{Taylor}}, \citenamefont {{Williams}}, \citenamefont
  {{Bullock}}, \citenamefont {{Dickinson}}, \citenamefont {{Finkelstein}},
  \citenamefont {{Natarajan}}, \citenamefont {{Richard}}, \citenamefont
  {{Robertson}}, \citenamefont {{Tumlinson}}, \citenamefont {{Zitrin}},
  \citenamefont {{Flanagan}}, \citenamefont {{Sembach}}, \citenamefont
  {{Soifer}},\ and\ \citenamefont {{Mountain}}}]{2017ApJ...837...97L}%
  \BibitemOpen
  \bibfield  {author} {\bibinfo {author} {\bibfnamefont {J.~M.}\ \bibnamefont
  {{Lotz}}}, \bibinfo {author} {\bibfnamefont {A.}~\bibnamefont {{Koekemoer}}},
  \bibinfo {author} {\bibfnamefont {D.}~\bibnamefont {{Coe}}}, \bibinfo
  {author} {\bibfnamefont {N.}~\bibnamefont {{Grogin}}}, \bibinfo {author}
  {\bibfnamefont {P.}~\bibnamefont {{Capak}}}, \bibinfo {author} {\bibfnamefont
  {J.}~\bibnamefont {{Mack}}}, \bibinfo {author} {\bibfnamefont
  {J.}~\bibnamefont {{Anderson}}}, \bibinfo {author} {\bibfnamefont
  {R.}~\bibnamefont {{Avila}}}, \bibinfo {author} {\bibfnamefont {E.~A.}\
  \bibnamefont {{Barker}}}, \bibinfo {author} {\bibfnamefont {D.}~\bibnamefont
  {{Borncamp}}}, \bibinfo {author} {\bibfnamefont {G.}~\bibnamefont
  {{Brammer}}}, \bibinfo {author} {\bibfnamefont {M.}~\bibnamefont {{Durbin}}},
  \bibinfo {author} {\bibfnamefont {H.}~\bibnamefont {{Gunning}}}, \bibinfo
  {author} {\bibfnamefont {B.}~\bibnamefont {{Hilbert}}}, \bibinfo {author}
  {\bibfnamefont {H.}~\bibnamefont {{Jenkner}}}, \bibinfo {author}
  {\bibfnamefont {H.}~\bibnamefont {{Khandrika}}}, \bibinfo {author}
  {\bibfnamefont {Z.}~\bibnamefont {{Levay}}}, \bibinfo {author} {\bibfnamefont
  {R.~A.}\ \bibnamefont {{Lucas}}}, \bibinfo {author} {\bibfnamefont
  {J.}~\bibnamefont {{MacKenty}}}, \bibinfo {author} {\bibfnamefont
  {S.}~\bibnamefont {{Ogaz}}}, \bibinfo {author} {\bibfnamefont
  {B.}~\bibnamefont {{Porterfield}}}, \bibinfo {author} {\bibfnamefont
  {N.}~\bibnamefont {{Reid}}}, \bibinfo {author} {\bibfnamefont
  {M.}~\bibnamefont {{Robberto}}}, \bibinfo {author} {\bibfnamefont
  {P.}~\bibnamefont {{Royle}}}, \bibinfo {author} {\bibfnamefont {L.~J.}\
  \bibnamefont {{Smith}}}, \bibinfo {author} {\bibfnamefont {L.~J.}\
  \bibnamefont {{Storrie-Lombardi}}}, \bibinfo {author} {\bibfnamefont
  {B.}~\bibnamefont {{Sunnquist}}}, \bibinfo {author} {\bibfnamefont
  {J.}~\bibnamefont {{Surace}}}, \bibinfo {author} {\bibfnamefont {D.~C.}\
  \bibnamefont {{Taylor}}}, \bibinfo {author} {\bibfnamefont {R.}~\bibnamefont
  {{Williams}}}, \bibinfo {author} {\bibfnamefont {J.}~\bibnamefont
  {{Bullock}}}, \bibinfo {author} {\bibfnamefont {M.}~\bibnamefont
  {{Dickinson}}}, \bibinfo {author} {\bibfnamefont {S.}~\bibnamefont
  {{Finkelstein}}}, \bibinfo {author} {\bibfnamefont {P.}~\bibnamefont
  {{Natarajan}}}, \bibinfo {author} {\bibfnamefont {J.}~\bibnamefont
  {{Richard}}}, \bibinfo {author} {\bibfnamefont {B.}~\bibnamefont
  {{Robertson}}}, \bibinfo {author} {\bibfnamefont {J.}~\bibnamefont
  {{Tumlinson}}}, \bibinfo {author} {\bibfnamefont {A.}~\bibnamefont
  {{Zitrin}}}, \bibinfo {author} {\bibfnamefont {K.}~\bibnamefont
  {{Flanagan}}}, \bibinfo {author} {\bibfnamefont {K.}~\bibnamefont
  {{Sembach}}}, \bibinfo {author} {\bibfnamefont {B.~T.}\ \bibnamefont
  {{Soifer}}}, \ and\ \bibinfo {author} {\bibfnamefont {M.}~\bibnamefont
  {{Mountain}}},\ }\href {\doibase 10.3847/1538-4357/837/1/97} {\bibfield
  {journal} {\bibinfo  {journal} {\apj}\ }\textbf {\bibinfo {volume} {837}},\
  \bibinfo {eid} {97} (\bibinfo {year} {2017})},\ \Eprint
  {http://arxiv.org/abs/1605.06567} {arXiv:1605.06567} \BibitemShut {NoStop}%
\bibitem [{\citenamefont {{Jauzac}}\ \emph {et~al.}(2015)\citenamefont
  {{Jauzac}}, \citenamefont {{Jullo}}, \citenamefont {{Eckert}}, \citenamefont
  {{Ebeling}}, \citenamefont {{Richard}}, \citenamefont {{Limousin}},
  \citenamefont {{Atek}}, \citenamefont {{Kneib}}, \citenamefont
  {{Cl{\'e}ment}}, \citenamefont {{Egami}}, \citenamefont {{Harvey}},
  \citenamefont {{Knowles}}, \citenamefont {{Massey}}, \citenamefont
  {{Natarajan}}, \citenamefont {{Neichel}},\ and\ \citenamefont
  {{Rexroth}}}]{2015MNRAS.446.4132J}%
  \BibitemOpen
  \bibfield  {author} {\bibinfo {author} {\bibfnamefont {M.}~\bibnamefont
  {{Jauzac}}}, \bibinfo {author} {\bibfnamefont {E.}~\bibnamefont {{Jullo}}},
  \bibinfo {author} {\bibfnamefont {D.}~\bibnamefont {{Eckert}}}, \bibinfo
  {author} {\bibfnamefont {H.}~\bibnamefont {{Ebeling}}}, \bibinfo {author}
  {\bibfnamefont {J.}~\bibnamefont {{Richard}}}, \bibinfo {author}
  {\bibfnamefont {M.}~\bibnamefont {{Limousin}}}, \bibinfo {author}
  {\bibfnamefont {H.}~\bibnamefont {{Atek}}}, \bibinfo {author} {\bibfnamefont
  {J.-P.}\ \bibnamefont {{Kneib}}}, \bibinfo {author} {\bibfnamefont
  {B.}~\bibnamefont {{Cl{\'e}ment}}}, \bibinfo {author} {\bibfnamefont
  {E.}~\bibnamefont {{Egami}}}, \bibinfo {author} {\bibfnamefont
  {D.}~\bibnamefont {{Harvey}}}, \bibinfo {author} {\bibfnamefont
  {K.}~\bibnamefont {{Knowles}}}, \bibinfo {author} {\bibfnamefont
  {R.}~\bibnamefont {{Massey}}}, \bibinfo {author} {\bibfnamefont
  {P.}~\bibnamefont {{Natarajan}}}, \bibinfo {author} {\bibfnamefont
  {B.}~\bibnamefont {{Neichel}}}, \ and\ \bibinfo {author} {\bibfnamefont
  {M.}~\bibnamefont {{Rexroth}}},\ }\href {\doibase 10.1093/mnras/stu2425}
  {\bibfield  {journal} {\bibinfo  {journal} {\mnras}\ }\textbf {\bibinfo
  {volume} {446}},\ \bibinfo {pages} {4132} (\bibinfo {year} {2015})},\ \Eprint
  {http://arxiv.org/abs/1406.3011} {arXiv:1406.3011} \BibitemShut {NoStop}%
\bibitem [{\citenamefont {{Karman}}\ \emph {et~al.}(2015)\citenamefont
  {{Karman}}, \citenamefont {{Caputi}}, \citenamefont {{Grillo}}, \citenamefont
  {{Balestra}}, \citenamefont {{Rosati}}, \citenamefont {{Vanzella}},
  \citenamefont {{Coe}}, \citenamefont {{Christensen}}, \citenamefont
  {{Koekemoer}}, \citenamefont {{Kr{\"u}hler}}, \citenamefont {{Lombardi}},
  \citenamefont {{Mercurio}}, \citenamefont {{Nonino}},\ and\ \citenamefont
  {{van der Wel}}}]{2015A&A...574A..11K}%
  \BibitemOpen
  \bibfield  {author} {\bibinfo {author} {\bibfnamefont {W.}~\bibnamefont
  {{Karman}}}, \bibinfo {author} {\bibfnamefont {K.~I.}\ \bibnamefont
  {{Caputi}}}, \bibinfo {author} {\bibfnamefont {C.}~\bibnamefont {{Grillo}}},
  \bibinfo {author} {\bibfnamefont {I.}~\bibnamefont {{Balestra}}}, \bibinfo
  {author} {\bibfnamefont {P.}~\bibnamefont {{Rosati}}}, \bibinfo {author}
  {\bibfnamefont {E.}~\bibnamefont {{Vanzella}}}, \bibinfo {author}
  {\bibfnamefont {D.}~\bibnamefont {{Coe}}}, \bibinfo {author} {\bibfnamefont
  {L.}~\bibnamefont {{Christensen}}}, \bibinfo {author} {\bibfnamefont {A.~M.}\
  \bibnamefont {{Koekemoer}}}, \bibinfo {author} {\bibfnamefont
  {T.}~\bibnamefont {{Kr{\"u}hler}}}, \bibinfo {author} {\bibfnamefont
  {M.}~\bibnamefont {{Lombardi}}}, \bibinfo {author} {\bibfnamefont
  {A.}~\bibnamefont {{Mercurio}}}, \bibinfo {author} {\bibfnamefont
  {M.}~\bibnamefont {{Nonino}}}, \ and\ \bibinfo {author} {\bibfnamefont
  {A.}~\bibnamefont {{van der Wel}}},\ }\href {\doibase
  10.1051/0004-6361/201424962} {\bibfield  {journal} {\bibinfo  {journal}
  {\aap}\ }\textbf {\bibinfo {volume} {574}},\ \bibinfo {eid} {A11} (\bibinfo
  {year} {2015})},\ \Eprint {http://arxiv.org/abs/1409.3507} {arXiv:1409.3507}
  \BibitemShut {NoStop}%
\bibitem [{\citenamefont {{Limousin}}\ \emph {et~al.}(2016)\citenamefont
  {{Limousin}}, \citenamefont {{Richard}}, \citenamefont {{Jullo}},
  \citenamefont {{Jauzac}}, \citenamefont {{Ebeling}}, \citenamefont
  {{Bonamigo}}, \citenamefont {{Alavi}}, \citenamefont {{Cl{\'e}ment}},
  \citenamefont {{Giocoli}}, \citenamefont {{Kneib}}, \citenamefont
  {{Verdugo}}, \citenamefont {{Natarajan}}, \citenamefont {{Siana}},
  \citenamefont {{Atek}},\ and\ \citenamefont
  {{Rexroth}}}]{2016A&A...588A..99L}%
  \BibitemOpen
  \bibfield  {author} {\bibinfo {author} {\bibfnamefont {M.}~\bibnamefont
  {{Limousin}}}, \bibinfo {author} {\bibfnamefont {J.}~\bibnamefont
  {{Richard}}}, \bibinfo {author} {\bibfnamefont {E.}~\bibnamefont {{Jullo}}},
  \bibinfo {author} {\bibfnamefont {M.}~\bibnamefont {{Jauzac}}}, \bibinfo
  {author} {\bibfnamefont {H.}~\bibnamefont {{Ebeling}}}, \bibinfo {author}
  {\bibfnamefont {M.}~\bibnamefont {{Bonamigo}}}, \bibinfo {author}
  {\bibfnamefont {A.}~\bibnamefont {{Alavi}}}, \bibinfo {author} {\bibfnamefont
  {B.}~\bibnamefont {{Cl{\'e}ment}}}, \bibinfo {author} {\bibfnamefont
  {C.}~\bibnamefont {{Giocoli}}}, \bibinfo {author} {\bibfnamefont {J.-P.}\
  \bibnamefont {{Kneib}}}, \bibinfo {author} {\bibfnamefont {T.}~\bibnamefont
  {{Verdugo}}}, \bibinfo {author} {\bibfnamefont {P.}~\bibnamefont
  {{Natarajan}}}, \bibinfo {author} {\bibfnamefont {B.}~\bibnamefont
  {{Siana}}}, \bibinfo {author} {\bibfnamefont {H.}~\bibnamefont {{Atek}}}, \
  and\ \bibinfo {author} {\bibfnamefont {M.}~\bibnamefont {{Rexroth}}},\ }\href
  {\doibase 10.1051/0004-6361/201527638} {\bibfield  {journal} {\bibinfo
  {journal} {\aap}\ }\textbf {\bibinfo {volume} {588}},\ \bibinfo {eid} {A99}
  (\bibinfo {year} {2016})},\ \Eprint {http://arxiv.org/abs/1510.08077}
  {arXiv:1510.08077} \BibitemShut {NoStop}%
\bibitem [{\citenamefont {{Lagattuta}}\ \emph {et~al.}(2017)\citenamefont
  {{Lagattuta}}, \citenamefont {{Richard}}, \citenamefont {{Cl{\'e}ment}},
  \citenamefont {{Mahler}}, \citenamefont {{Patr{\'{\i}}cio}}, \citenamefont
  {{Pell{\'o}}}, \citenamefont {{Soucail}}, \citenamefont {{Schmidt}},
  \citenamefont {{Wisotzki}}, \citenamefont {{Martinez}},\ and\ \citenamefont
  {{Bina}}}]{2017MNRAS.469.3946L}%
  \BibitemOpen
  \bibfield  {author} {\bibinfo {author} {\bibfnamefont {D.~J.}\ \bibnamefont
  {{Lagattuta}}}, \bibinfo {author} {\bibfnamefont {J.}~\bibnamefont
  {{Richard}}}, \bibinfo {author} {\bibfnamefont {B.}~\bibnamefont
  {{Cl{\'e}ment}}}, \bibinfo {author} {\bibfnamefont {G.}~\bibnamefont
  {{Mahler}}}, \bibinfo {author} {\bibfnamefont {V.}~\bibnamefont
  {{Patr{\'{\i}}cio}}}, \bibinfo {author} {\bibfnamefont {R.}~\bibnamefont
  {{Pell{\'o}}}}, \bibinfo {author} {\bibfnamefont {G.}~\bibnamefont
  {{Soucail}}}, \bibinfo {author} {\bibfnamefont {K.~B.}\ \bibnamefont
  {{Schmidt}}}, \bibinfo {author} {\bibfnamefont {L.}~\bibnamefont
  {{Wisotzki}}}, \bibinfo {author} {\bibfnamefont {J.}~\bibnamefont
  {{Martinez}}}, \ and\ \bibinfo {author} {\bibfnamefont {D.}~\bibnamefont
  {{Bina}}},\ }\href {\doibase 10.1093/mnras/stx1079} {\bibfield  {journal}
  {\bibinfo  {journal} {\mnras}\ }\textbf {\bibinfo {volume} {469}},\ \bibinfo
  {pages} {3946} (\bibinfo {year} {2017})},\ \Eprint
  {http://arxiv.org/abs/1611.01513} {arXiv:1611.01513} \BibitemShut {NoStop}%
\bibitem [{\citenamefont {{Mahler}}\ \emph {et~al.}(2018)\citenamefont
  {{Mahler}}, \citenamefont {{Richard}}, \citenamefont {{Cl{\'e}ment}},
  \citenamefont {{Lagattuta}}, \citenamefont {{Schmidt}}, \citenamefont
  {{Patr{\'{\i}}cio}}, \citenamefont {{Soucail}}, \citenamefont {{Bacon}},
  \citenamefont {{Pello}}, \citenamefont {{Bouwens}}, \citenamefont {{Maseda}},
  \citenamefont {{Martinez}}, \citenamefont {{Carollo}}, \citenamefont
  {{Inami}}, \citenamefont {{Leclercq}},\ and\ \citenamefont
  {{Wisotzki}}}]{2018MNRAS.473..663M}%
  \BibitemOpen
  \bibfield  {author} {\bibinfo {author} {\bibfnamefont {G.}~\bibnamefont
  {{Mahler}}}, \bibinfo {author} {\bibfnamefont {J.}~\bibnamefont {{Richard}}},
  \bibinfo {author} {\bibfnamefont {B.}~\bibnamefont {{Cl{\'e}ment}}}, \bibinfo
  {author} {\bibfnamefont {D.}~\bibnamefont {{Lagattuta}}}, \bibinfo {author}
  {\bibfnamefont {K.}~\bibnamefont {{Schmidt}}}, \bibinfo {author}
  {\bibfnamefont {V.}~\bibnamefont {{Patr{\'{\i}}cio}}}, \bibinfo {author}
  {\bibfnamefont {G.}~\bibnamefont {{Soucail}}}, \bibinfo {author}
  {\bibfnamefont {R.}~\bibnamefont {{Bacon}}}, \bibinfo {author} {\bibfnamefont
  {R.}~\bibnamefont {{Pello}}}, \bibinfo {author} {\bibfnamefont
  {R.}~\bibnamefont {{Bouwens}}}, \bibinfo {author} {\bibfnamefont
  {M.}~\bibnamefont {{Maseda}}}, \bibinfo {author} {\bibfnamefont
  {J.}~\bibnamefont {{Martinez}}}, \bibinfo {author} {\bibfnamefont
  {M.}~\bibnamefont {{Carollo}}}, \bibinfo {author} {\bibfnamefont
  {H.}~\bibnamefont {{Inami}}}, \bibinfo {author} {\bibfnamefont
  {F.}~\bibnamefont {{Leclercq}}}, \ and\ \bibinfo {author} {\bibfnamefont
  {L.}~\bibnamefont {{Wisotzki}}},\ }\href {\doibase 10.1093/mnras/stx1971}
  {\bibfield  {journal} {\bibinfo  {journal} {\mnras}\ }\textbf {\bibinfo
  {volume} {473}},\ \bibinfo {pages} {663} (\bibinfo {year} {2018})},\ \Eprint
  {http://arxiv.org/abs/1702.06962} {arXiv:1702.06962} \BibitemShut {NoStop}%
\bibitem [{\citenamefont {{Kawamata}}\ \emph {et~al.}(2018)\citenamefont
  {{Kawamata}}, \citenamefont {{Ishigaki}}, \citenamefont {{Shimasaku}},
  \citenamefont {{Oguri}}, \citenamefont {{Ouchi}},\ and\ \citenamefont
  {{Tanigawa}}}]{2018ApJ...855....4K}%
  \BibitemOpen
  \bibfield  {author} {\bibinfo {author} {\bibfnamefont {R.}~\bibnamefont
  {{Kawamata}}}, \bibinfo {author} {\bibfnamefont {M.}~\bibnamefont
  {{Ishigaki}}}, \bibinfo {author} {\bibfnamefont {K.}~\bibnamefont
  {{Shimasaku}}}, \bibinfo {author} {\bibfnamefont {M.}~\bibnamefont
  {{Oguri}}}, \bibinfo {author} {\bibfnamefont {M.}~\bibnamefont {{Ouchi}}}, \
  and\ \bibinfo {author} {\bibfnamefont {S.}~\bibnamefont {{Tanigawa}}},\
  }\href {\doibase 10.3847/1538-4357/aaa6cf} {\bibfield  {journal} {\bibinfo
  {journal} {\apj}\ }\textbf {\bibinfo {volume} {855}},\ \bibinfo {eid} {4}
  (\bibinfo {year} {2018})},\ \Eprint {http://arxiv.org/abs/1710.07301}
  {arXiv:1710.07301} \BibitemShut {NoStop}%
\bibitem [{\citenamefont {{Keeton}}(2001)}]{2001astro.ph..2340K}%
  \BibitemOpen
  \bibfield  {author} {\bibinfo {author} {\bibfnamefont {C.~R.}\ \bibnamefont
  {{Keeton}}},\ }\href@noop {} {\bibfield  {journal} {\bibinfo  {journal}
  {ArXiv Astrophysics e-prints}\ } (\bibinfo {year} {2001})},\ \Eprint
  {http://arxiv.org/abs/astro-ph/0102340} {astro-ph/0102340} \BibitemShut
  {NoStop}%
\bibitem [{\citenamefont {{Jullo}}\ \emph {et~al.}(2007)\citenamefont
  {{Jullo}}, \citenamefont {{Kneib}}, \citenamefont {{Limousin}}, \citenamefont
  {{El{\'{\i}}asd{\'o}ttir}}, \citenamefont {{Marshall}},\ and\ \citenamefont
  {{Verdugo}}}]{2007NJPh....9..447J}%
  \BibitemOpen
  \bibfield  {author} {\bibinfo {author} {\bibfnamefont {E.}~\bibnamefont
  {{Jullo}}}, \bibinfo {author} {\bibfnamefont {J.-P.}\ \bibnamefont
  {{Kneib}}}, \bibinfo {author} {\bibfnamefont {M.}~\bibnamefont {{Limousin}}},
  \bibinfo {author} {\bibfnamefont {{\'A}.}~\bibnamefont
  {{El{\'{\i}}asd{\'o}ttir}}}, \bibinfo {author} {\bibfnamefont {P.~J.}\
  \bibnamefont {{Marshall}}}, \ and\ \bibinfo {author} {\bibfnamefont
  {T.}~\bibnamefont {{Verdugo}}},\ }\href {\doibase 10.1088/1367-2630/9/12/447}
  {\bibfield  {journal} {\bibinfo  {journal} {New Journal of Physics}\ }\textbf
  {\bibinfo {volume} {9}},\ \bibinfo {pages} {447} (\bibinfo {year} {2007})},\
  \Eprint {http://arxiv.org/abs/0706.0048} {arXiv:0706.0048} \BibitemShut
  {NoStop}%
\bibitem [{\citenamefont {{Meneghetti}}\ \emph {et~al.}(2017)\citenamefont
  {{Meneghetti}}, \citenamefont {{Natarajan}}, \citenamefont {{Coe}},
  \citenamefont {{Contini}}, \citenamefont {{De Lucia}}, \citenamefont
  {{Giocoli}}, \citenamefont {{Acebron}}, \citenamefont {{Borgani}},
  \citenamefont {{Bradac}}, \citenamefont {{Diego}}, \citenamefont {{Hoag}},
  \citenamefont {{Ishigaki}}, \citenamefont {{Johnson}}, \citenamefont
  {{Jullo}}, \citenamefont {{Kawamata}}, \citenamefont {{Lam}}, \citenamefont
  {{Limousin}}, \citenamefont {{Liesenborgs}}, \citenamefont {{Oguri}},
  \citenamefont {{Sebesta}}, \citenamefont {{Sharon}}, \citenamefont
  {{Williams}},\ and\ \citenamefont {{Zitrin}}}]{2017MNRAS.472.3177M}%
  \BibitemOpen
  \bibfield  {author} {\bibinfo {author} {\bibfnamefont {M.}~\bibnamefont
  {{Meneghetti}}}, \bibinfo {author} {\bibfnamefont {P.}~\bibnamefont
  {{Natarajan}}}, \bibinfo {author} {\bibfnamefont {D.}~\bibnamefont {{Coe}}},
  \bibinfo {author} {\bibfnamefont {E.}~\bibnamefont {{Contini}}}, \bibinfo
  {author} {\bibfnamefont {G.}~\bibnamefont {{De Lucia}}}, \bibinfo {author}
  {\bibfnamefont {C.}~\bibnamefont {{Giocoli}}}, \bibinfo {author}
  {\bibfnamefont {A.}~\bibnamefont {{Acebron}}}, \bibinfo {author}
  {\bibfnamefont {S.}~\bibnamefont {{Borgani}}}, \bibinfo {author}
  {\bibfnamefont {M.}~\bibnamefont {{Bradac}}}, \bibinfo {author}
  {\bibfnamefont {J.~M.}\ \bibnamefont {{Diego}}}, \bibinfo {author}
  {\bibfnamefont {A.}~\bibnamefont {{Hoag}}}, \bibinfo {author} {\bibfnamefont
  {M.}~\bibnamefont {{Ishigaki}}}, \bibinfo {author} {\bibfnamefont {T.~L.}\
  \bibnamefont {{Johnson}}}, \bibinfo {author} {\bibfnamefont {E.}~\bibnamefont
  {{Jullo}}}, \bibinfo {author} {\bibfnamefont {R.}~\bibnamefont {{Kawamata}}},
  \bibinfo {author} {\bibfnamefont {D.}~\bibnamefont {{Lam}}}, \bibinfo
  {author} {\bibfnamefont {M.}~\bibnamefont {{Limousin}}}, \bibinfo {author}
  {\bibfnamefont {J.}~\bibnamefont {{Liesenborgs}}}, \bibinfo {author}
  {\bibfnamefont {M.}~\bibnamefont {{Oguri}}}, \bibinfo {author} {\bibfnamefont
  {K.}~\bibnamefont {{Sebesta}}}, \bibinfo {author} {\bibfnamefont
  {K.}~\bibnamefont {{Sharon}}}, \bibinfo {author} {\bibfnamefont {L.~L.~R.}\
  \bibnamefont {{Williams}}}, \ and\ \bibinfo {author} {\bibfnamefont
  {A.}~\bibnamefont {{Zitrin}}},\ }\href {\doibase 10.1093/mnras/stx2064}
  {\bibfield  {journal} {\bibinfo  {journal} {\mnras}\ }\textbf {\bibinfo
  {volume} {472}},\ \bibinfo {pages} {3177} (\bibinfo {year} {2017})},\ \Eprint
  {http://arxiv.org/abs/1606.04548} {arXiv:1606.04548} \BibitemShut {NoStop}%
\bibitem [{\citenamefont {{Petrushevska}}\ \emph {et~al.}(2018)\citenamefont
  {{Petrushevska}}, \citenamefont {{Goobar}}, \citenamefont {{Lagattuta}},
  \citenamefont {{Amanullah}}, \citenamefont {{Hangard}}, \citenamefont
  {{Fabbro}}, \citenamefont {{Lidman}}, \citenamefont {{Paech}}, \citenamefont
  {{Richard}},\ and\ \citenamefont {{Kneib}}}]{2018A&A...614A.103P}%
  \BibitemOpen
  \bibfield  {author} {\bibinfo {author} {\bibfnamefont {T.}~\bibnamefont
  {{Petrushevska}}}, \bibinfo {author} {\bibfnamefont {A.}~\bibnamefont
  {{Goobar}}}, \bibinfo {author} {\bibfnamefont {D.~J.}\ \bibnamefont
  {{Lagattuta}}}, \bibinfo {author} {\bibfnamefont {R.}~\bibnamefont
  {{Amanullah}}}, \bibinfo {author} {\bibfnamefont {L.}~\bibnamefont
  {{Hangard}}}, \bibinfo {author} {\bibfnamefont {S.}~\bibnamefont {{Fabbro}}},
  \bibinfo {author} {\bibfnamefont {C.}~\bibnamefont {{Lidman}}}, \bibinfo
  {author} {\bibfnamefont {K.}~\bibnamefont {{Paech}}}, \bibinfo {author}
  {\bibfnamefont {J.}~\bibnamefont {{Richard}}}, \ and\ \bibinfo {author}
  {\bibfnamefont {J.~P.}\ \bibnamefont {{Kneib}}},\ }\href {\doibase
  10.1051/0004-6361/201731552} {\bibfield  {journal} {\bibinfo  {journal}
  {\aap}\ }\textbf {\bibinfo {volume} {614}},\ \bibinfo {eid} {A103} (\bibinfo
  {year} {2018})},\ \Eprint {http://arxiv.org/abs/1802.10525}
  {arXiv:1802.10525} \BibitemShut {NoStop}%
\bibitem [{\citenamefont {{Kalirai}}(2018)}]{2018arXiv180506941K}%
  \BibitemOpen
  \bibfield  {author} {\bibinfo {author} {\bibfnamefont {J.}~\bibnamefont
  {{Kalirai}}},\ }\href@noop {} {\bibfield  {journal} {\bibinfo  {journal}
  {ArXiv e-prints}\ } (\bibinfo {year} {2018})},\ \Eprint
  {http://arxiv.org/abs/1805.06941} {arXiv:1805.06941 [astro-ph.IM]}
  \BibitemShut {NoStop}%
\bibitem [{\citenamefont {{Kim}}\ \emph {et~al.}(1996)\citenamefont {{Kim}},
  \citenamefont {{Goobar}},\ and\ \citenamefont
  {{Perlmutter}}}]{1996PASP..108..190K}%
  \BibitemOpen
  \bibfield  {author} {\bibinfo {author} {\bibfnamefont {A.}~\bibnamefont
  {{Kim}}}, \bibinfo {author} {\bibfnamefont {A.}~\bibnamefont {{Goobar}}}, \
  and\ \bibinfo {author} {\bibfnamefont {S.}~\bibnamefont {{Perlmutter}}},\
  }\href {\doibase 10.1086/133709} {\bibfield  {journal} {\bibinfo  {journal}
  {\pasp}\ }\textbf {\bibinfo {volume} {108}},\ \bibinfo {pages} {190}
  (\bibinfo {year} {1996})},\ \Eprint {http://arxiv.org/abs/astro-ph/9505024}
  {astro-ph/9505024} \BibitemShut {NoStop}%
\bibitem [{\citenamefont {{Kriek}}\ \emph {et~al.}(2009)\citenamefont
  {{Kriek}}, \citenamefont {{van Dokkum}}, \citenamefont {{Labb{\'e}}},
  \citenamefont {{Franx}}, \citenamefont {{Illingworth}}, \citenamefont
  {{Marchesini}},\ and\ \citenamefont {{Quadri}}}]{2009ApJ...700..221K}%
  \BibitemOpen
  \bibfield  {author} {\bibinfo {author} {\bibfnamefont {M.}~\bibnamefont
  {{Kriek}}}, \bibinfo {author} {\bibfnamefont {P.~G.}\ \bibnamefont {{van
  Dokkum}}}, \bibinfo {author} {\bibfnamefont {I.}~\bibnamefont {{Labb{\'e}}}},
  \bibinfo {author} {\bibfnamefont {M.}~\bibnamefont {{Franx}}}, \bibinfo
  {author} {\bibfnamefont {G.~D.}\ \bibnamefont {{Illingworth}}}, \bibinfo
  {author} {\bibfnamefont {D.}~\bibnamefont {{Marchesini}}}, \ and\ \bibinfo
  {author} {\bibfnamefont {R.~F.}\ \bibnamefont {{Quadri}}},\ }\href {\doibase
  10.1088/0004-637X/700/1/221} {\bibfield  {journal} {\bibinfo  {journal}
  {\apj}\ }\textbf {\bibinfo {volume} {700}},\ \bibinfo {pages} {221} (\bibinfo
  {year} {2009})},\ \Eprint {http://arxiv.org/abs/0905.1692} {arXiv:0905.1692
  [astro-ph.CO]} \BibitemShut {NoStop}%
\bibitem [{\citenamefont {{Scannapieco}}\ and\ \citenamefont
  {{Bildsten}}(2005)}]{2005ApJ...629L..85S}%
  \BibitemOpen
  \bibfield  {author} {\bibinfo {author} {\bibfnamefont {E.}~\bibnamefont
  {{Scannapieco}}}\ and\ \bibinfo {author} {\bibfnamefont {L.}~\bibnamefont
  {{Bildsten}}},\ }\href {\doibase 10.1086/452632} {\bibfield  {journal}
  {\bibinfo  {journal} {\apjl}\ }\textbf {\bibinfo {volume} {629}},\ \bibinfo
  {pages} {L85} (\bibinfo {year} {2005})},\ \Eprint
  {http://arxiv.org/abs/astro-ph/0507456} {astro-ph/0507456} \BibitemShut
  {NoStop}%
\bibitem [{\citenamefont {{Smith}}\ \emph {et~al.}(2012)\citenamefont
  {{Smith}}, \citenamefont {{Nichol}}, \citenamefont {{Dilday}}, \citenamefont
  {{Marriner}}, \citenamefont {{Kessler}}, \citenamefont {{Bassett}},
  \citenamefont {{Cinabro}}, \citenamefont {{Frieman}}, \citenamefont
  {{Garnavich}}, \citenamefont {{Jha}}, \citenamefont {{Lampeitl}},
  \citenamefont {{Sako}}, \citenamefont {{Schneider}},\ and\ \citenamefont
  {{Sollerman}}}]{2012ApJ...755...61S}%
  \BibitemOpen
  \bibfield  {author} {\bibinfo {author} {\bibfnamefont {M.}~\bibnamefont
  {{Smith}}}, \bibinfo {author} {\bibfnamefont {R.~C.}\ \bibnamefont
  {{Nichol}}}, \bibinfo {author} {\bibfnamefont {B.}~\bibnamefont {{Dilday}}},
  \bibinfo {author} {\bibfnamefont {J.}~\bibnamefont {{Marriner}}}, \bibinfo
  {author} {\bibfnamefont {R.}~\bibnamefont {{Kessler}}}, \bibinfo {author}
  {\bibfnamefont {B.}~\bibnamefont {{Bassett}}}, \bibinfo {author}
  {\bibfnamefont {D.}~\bibnamefont {{Cinabro}}}, \bibinfo {author}
  {\bibfnamefont {J.}~\bibnamefont {{Frieman}}}, \bibinfo {author}
  {\bibfnamefont {P.}~\bibnamefont {{Garnavich}}}, \bibinfo {author}
  {\bibfnamefont {S.~W.}\ \bibnamefont {{Jha}}}, \bibinfo {author}
  {\bibfnamefont {H.}~\bibnamefont {{Lampeitl}}}, \bibinfo {author}
  {\bibfnamefont {M.}~\bibnamefont {{Sako}}}, \bibinfo {author} {\bibfnamefont
  {D.~P.}\ \bibnamefont {{Schneider}}}, \ and\ \bibinfo {author} {\bibfnamefont
  {J.}~\bibnamefont {{Sollerman}}},\ }\href {\doibase
  10.1088/0004-637X/755/1/61} {\bibfield  {journal} {\bibinfo  {journal}
  {\apj}\ }\textbf {\bibinfo {volume} {755}},\ \bibinfo {eid} {61} (\bibinfo
  {year} {2012})},\ \Eprint {http://arxiv.org/abs/1108.4923} {arXiv:1108.4923}
  \BibitemShut {NoStop}%
\bibitem [{\citenamefont {{Goobar}}\ \emph {et~al.}(2009)\citenamefont
  {{Goobar}}, \citenamefont {{Paech}}, \citenamefont {{Stanishev}},
  \citenamefont {{Amanullah}}, \citenamefont {{Dahl{\'e}n}}, \citenamefont
  {{J{\"o}nsson}}, \citenamefont {{Kneib}}, \citenamefont {{Lidman}},
  \citenamefont {{Limousin}}, \citenamefont {{M{\"o}rtsell}}, \citenamefont
  {{Nobili}}, \citenamefont {{Richard}}, \citenamefont {{Riehm}},\ and\
  \citenamefont {{von Strauss}}}]{Second}%
  \BibitemOpen
  \bibfield  {author} {\bibinfo {author} {\bibfnamefont {A.}~\bibnamefont
  {{Goobar}}}, \bibinfo {author} {\bibfnamefont {K.}~\bibnamefont {{Paech}}},
  \bibinfo {author} {\bibfnamefont {V.}~\bibnamefont {{Stanishev}}}, \bibinfo
  {author} {\bibfnamefont {R.}~\bibnamefont {{Amanullah}}}, \bibinfo {author}
  {\bibfnamefont {T.}~\bibnamefont {{Dahl{\'e}n}}}, \bibinfo {author}
  {\bibfnamefont {J.}~\bibnamefont {{J{\"o}nsson}}}, \bibinfo {author}
  {\bibfnamefont {J.~P.}\ \bibnamefont {{Kneib}}}, \bibinfo {author}
  {\bibfnamefont {C.}~\bibnamefont {{Lidman}}}, \bibinfo {author}
  {\bibfnamefont {M.}~\bibnamefont {{Limousin}}}, \bibinfo {author}
  {\bibfnamefont {E.}~\bibnamefont {{M{\"o}rtsell}}}, \bibinfo {author}
  {\bibfnamefont {S.}~\bibnamefont {{Nobili}}}, \bibinfo {author}
  {\bibfnamefont {J.}~\bibnamefont {{Richard}}}, \bibinfo {author}
  {\bibfnamefont {T.}~\bibnamefont {{Riehm}}}, \ and\ \bibinfo {author}
  {\bibfnamefont {M.}~\bibnamefont {{von Strauss}}},\ }\href {\doibase
  10.1051/0004-6361/200811254} {\bibfield  {journal} {\bibinfo  {journal}
  {\aap}\ }\textbf {\bibinfo {volume} {507}},\ \bibinfo {pages} {71} (\bibinfo
  {year} {2009})},\ \Eprint {http://arxiv.org/abs/0810.4932} {arXiv:0810.4932}
  \BibitemShut {NoStop}%
\bibitem [{\citenamefont {{Beichman}}\ \emph {et~al.}(2012)\citenamefont
  {{Beichman}}, \citenamefont {{Rieke}}, \citenamefont {{Eisenstein}},
  \citenamefont {{Greene}}, \citenamefont {{Krist}}, \citenamefont
  {{McCarthy}}, \citenamefont {{Meyer}},\ and\ \citenamefont
  {{Stansberry}}}]{2012SPIE.8442E..2NB}%
  \BibitemOpen
  \bibfield  {author} {\bibinfo {author} {\bibfnamefont {C.~A.}\ \bibnamefont
  {{Beichman}}}, \bibinfo {author} {\bibfnamefont {M.}~\bibnamefont {{Rieke}}},
  \bibinfo {author} {\bibfnamefont {D.}~\bibnamefont {{Eisenstein}}}, \bibinfo
  {author} {\bibfnamefont {T.~P.}\ \bibnamefont {{Greene}}}, \bibinfo {author}
  {\bibfnamefont {J.}~\bibnamefont {{Krist}}}, \bibinfo {author} {\bibfnamefont
  {D.}~\bibnamefont {{McCarthy}}}, \bibinfo {author} {\bibfnamefont
  {M.}~\bibnamefont {{Meyer}}}, \ and\ \bibinfo {author} {\bibfnamefont
  {J.}~\bibnamefont {{Stansberry}}},\ }in\ \href {\doibase 10.1117/12.925447}
  {\emph {\bibinfo {booktitle} {Space Telescopes and Instrumentation 2012:
  Optical, Infrared, and Millimeter Wave}}},\ \bibinfo {series} {\procspie},
  Vol.\ \bibinfo {volume} {8442}\ (\bibinfo {year} {2012})\ p.\ \bibinfo
  {pages} {84422N}\BibitemShut {NoStop}%
\bibitem [{\citenamefont {{Bina}}\ \emph {et~al.}(2016)\citenamefont {{Bina}},
  \citenamefont {{Pell{\'o}}}, \citenamefont {{Richard}}, \citenamefont
  {{Lewis}}, \citenamefont {{Patr{\'{\i}}cio}}, \citenamefont {{Cantalupo}},
  \citenamefont {{Herenz}}, \citenamefont {{Soto}}, \citenamefont
  {{Weilbacher}}, \citenamefont {{Bacon}}, \citenamefont {{Vernet}},
  \citenamefont {{Wisotzki}}, \citenamefont {{Cl{\'e}ment}}, \citenamefont
  {{Cuby}}, \citenamefont {{Lagattuta}}, \citenamefont {{Soucail}},\ and\
  \citenamefont {{Verhamme}}}]{2016A&A...590A..14B}%
  \BibitemOpen
  \bibfield  {author} {\bibinfo {author} {\bibfnamefont {D.}~\bibnamefont
  {{Bina}}}, \bibinfo {author} {\bibfnamefont {R.}~\bibnamefont {{Pell{\'o}}}},
  \bibinfo {author} {\bibfnamefont {J.}~\bibnamefont {{Richard}}}, \bibinfo
  {author} {\bibfnamefont {J.}~\bibnamefont {{Lewis}}}, \bibinfo {author}
  {\bibfnamefont {V.}~\bibnamefont {{Patr{\'{\i}}cio}}}, \bibinfo {author}
  {\bibfnamefont {S.}~\bibnamefont {{Cantalupo}}}, \bibinfo {author}
  {\bibfnamefont {E.~C.}\ \bibnamefont {{Herenz}}}, \bibinfo {author}
  {\bibfnamefont {K.}~\bibnamefont {{Soto}}}, \bibinfo {author} {\bibfnamefont
  {P.~M.}\ \bibnamefont {{Weilbacher}}}, \bibinfo {author} {\bibfnamefont
  {R.}~\bibnamefont {{Bacon}}}, \bibinfo {author} {\bibfnamefont {J.~D.~R.}\
  \bibnamefont {{Vernet}}}, \bibinfo {author} {\bibfnamefont {L.}~\bibnamefont
  {{Wisotzki}}}, \bibinfo {author} {\bibfnamefont {B.}~\bibnamefont
  {{Cl{\'e}ment}}}, \bibinfo {author} {\bibfnamefont {J.~G.}\ \bibnamefont
  {{Cuby}}}, \bibinfo {author} {\bibfnamefont {D.~J.}\ \bibnamefont
  {{Lagattuta}}}, \bibinfo {author} {\bibfnamefont {G.}~\bibnamefont
  {{Soucail}}}, \ and\ \bibinfo {author} {\bibfnamefont {A.}~\bibnamefont
  {{Verhamme}}},\ }\href {\doibase 10.1051/0004-6361/201527913} {\bibfield
  {journal} {\bibinfo  {journal} {\aap}\ }\textbf {\bibinfo {volume} {590}},\
  \bibinfo {eid} {A14} (\bibinfo {year} {2016})},\ \Eprint
  {http://arxiv.org/abs/1603.05833} {arXiv:1603.05833} \BibitemShut {NoStop}%
\bibitem [{\citenamefont {{Goobar}}\ \emph {et~al.}(2017)\citenamefont
  {{Goobar}}, \citenamefont {{Amanullah}}, \citenamefont {{Kulkarni}},
  \citenamefont {{Nugent}}, \citenamefont {{Johansson}}, \citenamefont
  {{Steidel}}, \citenamefont {{Law}}, \citenamefont {{M{\"o}rtsell}},
  \citenamefont {{Quimby}}, \citenamefont {{Blagorodnova}}, \citenamefont
  {{Brandeker}}, \citenamefont {{Cao}}, \citenamefont {{Cooray}}, \citenamefont
  {{Ferretti}}, \citenamefont {{Fremling}}, \citenamefont {{Hangard}},
  \citenamefont {{Kasliwal}}, \citenamefont {{Kupfer}}, \citenamefont
  {{Lunnan}}, \citenamefont {{Masci}}, \citenamefont {{Miller}}, \citenamefont
  {{Nayyeri}}, \citenamefont {{Neill}}, \citenamefont {{Ofek}}, \citenamefont
  {{Papadogiannakis}}, \citenamefont {{Petrushevska}}, \citenamefont {{Ravi}},
  \citenamefont {{Sollerman}}, \citenamefont {{Sullivan}}, \citenamefont
  {{Taddia}}, \citenamefont {{Walters}}, \citenamefont {{Wilson}},
  \citenamefont {{Yan}},\ and\ \citenamefont {{Yaron}}}]{2017Sci...356..291G}%
  \BibitemOpen
  \bibfield  {author} {\bibinfo {author} {\bibfnamefont {A.}~\bibnamefont
  {{Goobar}}}, \bibinfo {author} {\bibfnamefont {R.}~\bibnamefont
  {{Amanullah}}}, \bibinfo {author} {\bibfnamefont {S.~R.}\ \bibnamefont
  {{Kulkarni}}}, \bibinfo {author} {\bibfnamefont {P.~E.}\ \bibnamefont
  {{Nugent}}}, \bibinfo {author} {\bibfnamefont {J.}~\bibnamefont
  {{Johansson}}}, \bibinfo {author} {\bibfnamefont {C.}~\bibnamefont
  {{Steidel}}}, \bibinfo {author} {\bibfnamefont {D.}~\bibnamefont {{Law}}},
  \bibinfo {author} {\bibfnamefont {E.}~\bibnamefont {{M{\"o}rtsell}}},
  \bibinfo {author} {\bibfnamefont {R.}~\bibnamefont {{Quimby}}}, \bibinfo
  {author} {\bibfnamefont {N.}~\bibnamefont {{Blagorodnova}}}, \bibinfo
  {author} {\bibfnamefont {A.}~\bibnamefont {{Brandeker}}}, \bibinfo {author}
  {\bibfnamefont {Y.}~\bibnamefont {{Cao}}}, \bibinfo {author} {\bibfnamefont
  {A.}~\bibnamefont {{Cooray}}}, \bibinfo {author} {\bibfnamefont
  {R.}~\bibnamefont {{Ferretti}}}, \bibinfo {author} {\bibfnamefont
  {C.}~\bibnamefont {{Fremling}}}, \bibinfo {author} {\bibfnamefont
  {L.}~\bibnamefont {{Hangard}}}, \bibinfo {author} {\bibfnamefont
  {M.}~\bibnamefont {{Kasliwal}}}, \bibinfo {author} {\bibfnamefont
  {T.}~\bibnamefont {{Kupfer}}}, \bibinfo {author} {\bibfnamefont
  {R.}~\bibnamefont {{Lunnan}}}, \bibinfo {author} {\bibfnamefont
  {F.}~\bibnamefont {{Masci}}}, \bibinfo {author} {\bibfnamefont {A.~A.}\
  \bibnamefont {{Miller}}}, \bibinfo {author} {\bibfnamefont {H.}~\bibnamefont
  {{Nayyeri}}}, \bibinfo {author} {\bibfnamefont {J.~D.}\ \bibnamefont
  {{Neill}}}, \bibinfo {author} {\bibfnamefont {E.~O.}\ \bibnamefont {{Ofek}}},
  \bibinfo {author} {\bibfnamefont {S.}~\bibnamefont {{Papadogiannakis}}},
  \bibinfo {author} {\bibfnamefont {T.}~\bibnamefont {{Petrushevska}}},
  \bibinfo {author} {\bibfnamefont {V.}~\bibnamefont {{Ravi}}}, \bibinfo
  {author} {\bibfnamefont {J.}~\bibnamefont {{Sollerman}}}, \bibinfo {author}
  {\bibfnamefont {M.}~\bibnamefont {{Sullivan}}}, \bibinfo {author}
  {\bibfnamefont {F.}~\bibnamefont {{Taddia}}}, \bibinfo {author}
  {\bibfnamefont {R.}~\bibnamefont {{Walters}}}, \bibinfo {author}
  {\bibfnamefont {D.}~\bibnamefont {{Wilson}}}, \bibinfo {author}
  {\bibfnamefont {L.}~\bibnamefont {{Yan}}}, \ and\ \bibinfo {author}
  {\bibfnamefont {O.}~\bibnamefont {{Yaron}}},\ }\href {\doibase
  10.1126/science.aal2729} {\bibfield  {journal} {\bibinfo  {journal}
  {Science}\ }\textbf {\bibinfo {volume} {356}},\ \bibinfo {pages} {291}
  (\bibinfo {year} {2017})},\ \Eprint {http://arxiv.org/abs/1611.00014}
  {arXiv:1611.00014} \BibitemShut {NoStop}%
\bibitem [{\citenamefont {{Goldstein}}\ \emph {et~al.}(2018)\citenamefont
  {{Goldstein}}, \citenamefont {{Nugent}}, \citenamefont {{Kasen}},\ and\
  \citenamefont {{Collett}}}]{2018ApJ...855...22G}%
  \BibitemOpen
  \bibfield  {author} {\bibinfo {author} {\bibfnamefont {D.~A.}\ \bibnamefont
  {{Goldstein}}}, \bibinfo {author} {\bibfnamefont {P.~E.}\ \bibnamefont
  {{Nugent}}}, \bibinfo {author} {\bibfnamefont {D.~N.}\ \bibnamefont
  {{Kasen}}}, \ and\ \bibinfo {author} {\bibfnamefont {T.~E.}\ \bibnamefont
  {{Collett}}},\ }\href {\doibase 10.3847/1538-4357/aaa975} {\bibfield
  {journal} {\bibinfo  {journal} {\apj}\ }\textbf {\bibinfo {volume} {855}},\
  \bibinfo {eid} {22} (\bibinfo {year} {2018})},\ \Eprint
  {http://arxiv.org/abs/1708.00003} {arXiv:1708.00003} \BibitemShut {NoStop}%
\bibitem [{\citenamefont {{Gilmozzi}}\ and\ \citenamefont
  {{Spyromilio}}(2007)}]{2007Msngr.127...11G}%
  \BibitemOpen
  \bibfield  {author} {\bibinfo {author} {\bibfnamefont {R.}~\bibnamefont
  {{Gilmozzi}}}\ and\ \bibinfo {author} {\bibfnamefont {J.}~\bibnamefont
  {{Spyromilio}}},\ }\href@noop {} {\bibfield  {journal} {\bibinfo  {journal}
  {The Messenger}\ }\textbf {\bibinfo {volume} {127}} (\bibinfo {year}
  {2007})}\BibitemShut {NoStop}%
\bibitem [{\citenamefont {{Johns}}\ \emph {et~al.}(2012)\citenamefont
  {{Johns}}, \citenamefont {{McCarthy}}, \citenamefont {{Raybould}},
  \citenamefont {{Bouchez}}, \citenamefont {{Farahani}}, \citenamefont
  {{Filgueira}}, \citenamefont {{Jacoby}}, \citenamefont {{Shectman}},\ and\
  \citenamefont {{Sheehan}}}]{2012SPIE.8444E..1HJ}%
  \BibitemOpen
  \bibfield  {author} {\bibinfo {author} {\bibfnamefont {M.}~\bibnamefont
  {{Johns}}}, \bibinfo {author} {\bibfnamefont {P.}~\bibnamefont {{McCarthy}}},
  \bibinfo {author} {\bibfnamefont {K.}~\bibnamefont {{Raybould}}}, \bibinfo
  {author} {\bibfnamefont {A.}~\bibnamefont {{Bouchez}}}, \bibinfo {author}
  {\bibfnamefont {A.}~\bibnamefont {{Farahani}}}, \bibinfo {author}
  {\bibfnamefont {J.}~\bibnamefont {{Filgueira}}}, \bibinfo {author}
  {\bibfnamefont {G.}~\bibnamefont {{Jacoby}}}, \bibinfo {author}
  {\bibfnamefont {S.}~\bibnamefont {{Shectman}}}, \ and\ \bibinfo {author}
  {\bibfnamefont {M.}~\bibnamefont {{Sheehan}}},\ }in\ \href {\doibase
  10.1117/12.926716} {\emph {\bibinfo {booktitle} {Ground-based and Airborne
  Telescopes IV}}},\ \bibinfo {series} {\procspie}, Vol.\ \bibinfo {volume}
  {8444}\ (\bibinfo {year} {2012})\ p.\ \bibinfo {pages} {84441H}\BibitemShut
  {NoStop}%
\end{thebibliography}%
\appendix

\section{}\label{App}
\begin{longrotatetable}
	\begin{deluxetable}{c c c c c c c c c c c c c c}
		\tabletypesize{\small}
		\tablecolumns{12}
		\tablewidth{5\textwidth}
		\centering
		\tablecaption{roperties of the multiply-imaged galaxies behind five HFF galaxy clusters. Time delays ($\Delta t$) of the multiply-imaged galaxies as predicted from the lensing model in K18 are listed in column 3. The time delays are measured relative to the first image in column 1.  The reference image in each case is the one with the shortest path length between the the galaxy and us. The time delays are with respect to the reference image. The magnification is given in magnitudes in column 4 and 5, and is computed using the lensing model in K18. Estimates of the core-collapse and SN Ia rates are given with $1\sigma$ errors in column 6 and 7, respectively. Given that the star formation rate of the galaxy depends on the luminosity of the galaxy, it is dependent of the predicted magnification, thus the lensing model. In columns 8-14, the detectability of a lensed SN for the different SN types assuming a 27.5 mag in F150W, e.g. in a survey using 1 hour exposure of the JWST/NIRCam instrument. In order to be observable, the luminosity of both images must lie above the observation threshold. \label{table:multi}}
		\tablehead{
			\colhead{Images} & \colhead{$z$ } & \colhead{$\Delta t$} & \colhead{$\Delta m_1$} & \colhead{$\Delta m_2$.}   & \colhead{$R_{\rm CC}$ $\times$  $10^{2}$} & \colhead{$R_{\rm Ia}$ $\times$ $10^{3}$} &  \multicolumn{7}{c}{Detectable SN types}\\			
			\colhead{ } & \colhead{} & \colhead{d} & \colhead{mag} & \colhead{mag}   & \colhead{yr$^{-1}$} & \colhead{yr$^{-1}$} & Ia & IIP & IIL & IIn & Ib & Ic & faint\tablenotemark{a}\\	}
\startdata
\hline
    \multicolumn{13}{c}{A2744}\\ 
\hline
1.3 + 1.1 & 1.688 & 2810(100) & 1.52(0.04) & 1.86(0.04) & 1.84(0.42) & 1.17(0.24) & y & y &  y&  y &  y &  y &  y \\ 
1.3 + 1.2 & $\cdots$  & 3120(100) &  $\cdots$ & 1.86(0.06) &  $\cdots$ &  $\cdots$ & y & y &  y&  y &  y &  y &  y \\ 
2.2 + 2.1 & 1.888 & 8450(170) & 1.10(0.04) & 2.31(0.07) & 0.12(0.09) & 0.07(0.05) & y & y &  y&  y &  y &  y &    \\ 
2.2 + 2.4 & $\cdots$  & 8780(170) &  $\cdots$ & 2.03(0.09) &  $\cdots$ &  $\cdots$ & y & y &  y&  y &  y &  y &    \\ 
3.3 + 3.1 & 3.980 & 6030(160) & 1.40(0.05) & 3.79(0.17) & 0.41(0.20) & 0.23(0.11) & y &   &   &  y &    &    &    \\ 
3.3 + 3.2 & $\cdots$  & 6040(170) &  $\cdots$ & 3.75(0.12) &  $\cdots$ &  $\cdots$ & y &   &   &  y &    &    &    \\ 
4.3 + 4.1 & 3.580 & 12340(230) & 2.18(0.26) & 1.69(0.06) & 4.46(3.58) & 2.92(2.39) & y &   &  y&  y &    &    &    \\ 
4.3 + 4.2 & $\cdots$  & 12010(220) &  $\cdots$ & 2.34(0.06) &  $\cdots$ &  $\cdots$ & y &   &  y&  y &  y &    &    \\ 
4.3 + 4.4 & $\cdots$ & 12680(240) & $\cdots$ & 2.33(0.29) & $\cdots$ & $\cdots$ & y &   &  y&  y &  y &    &    \\ 
4.3 + 4.5 & $\cdots$  & 12680(240) &  $\cdots$ & 2.55(0.17) &  $\cdots$ &  $\cdots$ & y &   &  y&  y &  y &    & \\ 
6.1 + 6.2 & 2.019 & 2700(90) & 1.35(0.04) & 1.37(0.05) & 0.53(0.03) & 0.30(0.02) & y & y &  y&  y &  y &  y &    \\ 
6.1 + 6.3 & $\cdots$  & 1240(90) &  $\cdots$ & 1.78(0.05) &  $\cdots$ &  $\cdots$ & y & y &  y&  y &  y &  y &    \\ 
8.3 + 8.1 & 3.975 & 7370(180) & 1.36(0.05) & 3.53(0.14) & 0.14(0.03) & 0.08(0.01) & y &   &   &  y &    &    &    \\ 
8.3 + 8.2 & $\cdots$  & 7370(180) &  $\cdots$ & 3.35(0.08) &  $\cdots$ &  $\cdots$ & y &   &   &  y &    &    &    \\ 
10.3 + 10.1 & 2.656 & 12280(220) & 1.24(0.05) & 4.55(0.19) & 0.016(0.003) & 0.009(0.002)  & y & y &  y&  y &  y &  y &    \\ 
10.3 + 10.2 & $\cdots$  & 12270(220) &  $\cdots$ & 4.66(0.21) &  $\cdots$ &  $\cdots$ & y & y &  y&  y &  y &  y &    \\ 
24.3 + 24.1 & 1.043 & 2630(80) & 1.51(0.03) & 2.34(0.03) & 0.014(0.004) & 0.008(0.002) & y & y &  y&  y &  y &  y &  y \\ 
24.3 + 24.2 & $\cdots$  & 2690(90) &  $\cdots$ & 2.03(0.07) &  $\cdots$ &  $\cdots$ & y & y &  y&  y &  y &  y &  y \\ 
26.3 + 26.1 & 3.054 & 2430(120) & 1.35(0.04) & 2.37(0.06) & 0.11(0.10) & 0.06(0.06) & y & y &  y&  y &  y &    &    \\ 
26.3 + 26.2 & $\cdots$  & 2320(120) &  $\cdots$ & 2.70(0.06) &  $\cdots$ &  $\cdots$ & y & y &  y&  y &  y &    &    \\ 
30.3 + 30.1 & 1.025 & 1380(80) & 1.59(0.04) & 1.86(0.06) & 0.04(0.01) & 0.02(0.01) & y & y &  y&  y &  y &  y &  y \\ 
30.3 + 30.2 & $\cdots$  & 1970(80) &  $\cdots$ & 2.05(0.11) &  $\cdots$ &  $\cdots$ & y & y &  y&  y &  y &  y &  y \\ 
34.3 + 34.1 & 3.785 & 1260(130) & 1.57(0.04) & 4.87(0.85) & 1.65(1.63) & 0.93(0.91) & y &   &  y&  y &    &    &    \\ 
34.3 + 34.2 & $\cdots$  & 1260(130) &  $\cdots$ & 5.12(0.47) &  $\cdots$ &  $\cdots$ & y &   &  y&  y &    &    &    \\ 
42.3 + 42.1 & 3.692 & 5690(150) & 1.43(0.05) & 1.58(0.05) & 0.38(0.25) & 0.21(0.14) & y &   &  y&  y &    &    &    \\ 
42.3 + 42.2 & $\cdots$  & 7980(170) &  $\cdots$ & 0.87(0.06) &  $\cdots$ &  $\cdots$ & y &   &  y&  y &    &    &    \\ 
42.3 + 42.4 & $\cdots$  & 7990(180) & 1.43(0.05) & 0.88(0.06) & $\cdots$  & $\cdots$  & y &   &  y&  y &    &    &    \\ 
42.3 + 42.5 & $\cdots$  & 8320(220) &  $\cdots$ & -0.76(0.19) &  $\cdots$ &  $\cdots$ & y &   &   &  y &    &    &    \\
\hline
    \multicolumn{13}{c}{AS1063}\\ 
\hline
1.3 + 1.1 & 1.228 & 2080(70) & 1.78(0.02) & 2.84(0.06) & 1.29(0.68) & 0.76(0.40) & y & y &  y&  y &  y &  y &  y \\ 
1.3 + 1.2 & $\cdots$  & 2030(70) &  $\cdots$ & 2.95(0.07) &  $\cdots$ &  $\cdots$ & y & y &  y&  y &  y &  y &  y \\ 
2.3 + 2.1 & 1.259 & 8930(120) & 1.23(0.02) & 3.12(0.05) & 0.47(0.25) & 0.29(0.15) & y & y &  y&  y &  y &  y &  y \\ 
2.3 + 2.2 & $\cdots$  & 8950(120) &  $\cdots$ & 3.07(0.05) &  $\cdots$ &  $\cdots$ & y & y &  y&  y &  y &  y &  y \\ 
4.3 + 4.1 & 1.258 & 6470(100) & 1.35(0.02) & 3.70(0.08) & 0.008(0.002) & 0.005(0.001) & y & y &  y&  y &  y &  y &  y \\ 
4.3 + 4.2 & $\cdots$  & 6470(100) &  $\cdots$ & 3.67(0.08) &  $\cdots$ &  $\cdots$ & y & y &  y&  y &  y &  y &  y \\ 
5.3 + 5.1 & 1.397 & 4450(110) & 1.31(0.03) & 1.64(0.03) & 4.42(1.75) & 2.55(0.96) & y & y &  y&  y &  y &  y &  y \\ 
5.3 + 5.2 & $\cdots$  & 7790(100) &  $\cdots$ & 1.31(0.04) &  $\cdots$ &  $\cdots$ & y & y &  y&  y &  y &  y &  y \\ 
6.3 + 6.1 & 1.428 & 9900(150) & 1.35(0.03) & 2.12(0.04) & 4.52(2.11) & 2.70(1.18) & y & y &  y&  y &  y &  y &  y \\ 
6.3 + 6.2 & $\cdots$  & 10510(160) &  $\cdots$ & 2.07(0.06) &  $\cdots$ &  $\cdots$ & y & y &  y&  y &  y &  y &  y \\ 
7.1 + 7.2 & 1.837 & 4860(200) & 1.43(0.03) & 1.79(0.04) & 2.21(1.38) & 1.25(0.76) & y & y &  y&  y &  y &  y &    \\ 
7.1 + 7.3 & $\cdots$  & 3570(160) &  $\cdots$ & 2.15(0.04) &  $\cdots$ &  $\cdots$ & y & y &  y&  y &  y &  y &    \\ 
11.3 + 11.1 & 3.117 & 3350(180) & 1.61(0.04) & 3.43(0.08) & 0.16(0.02) & 0.09(0.01) & y & y &  y&  y &  y &  y &    \\ 
11.3 + 11.2 & $\cdots$  & 3370(170) &  $\cdots$ & 3.31(0.06) &  $\cdots$ &  $\cdots$ & y & y &  y&  y &  y &  y &    \\ 
14.3 + 14.1 & 3.118 & 19540(290) & 1.04(0.03) & 1.80(0.14) & 0.51(0.04) & 0.29(0.02) & y &   &  y&  y &  y &    &    \\ 
14.3 + 14.2 & $\cdots$  & 17640(250) &  $\cdots$ & 1.95(0.04) &  $\cdots$ &  $\cdots$ & y &   &  y&  y &  y &    &    \\ 
15.3 + 15.1 & 3.060 & 25110(270) & 0.96(0.02) & 2.05(0.04) & 0.34(0.03) & 0.19(0.01) & y &   &  y&  y &  y &    &    \\ 
15.3 + 15.2 & $\cdots$  & 24710(250) &  $\cdots$ & 2.42(0.04) &  $\cdots$ &  $\cdots$ & y &   &  y&  y &  y &    &    \\ 
17.3 + 17.1 & 3.606 & 21380(280) & 1.00(0.03) & 2.24(0.05) & 2.02(1.82) & 1.16(1.01) & y &   &  y&  y &    &    &    \\ 
17.3 + 17.2 & $\cdots$  & 22170(290) &  $\cdots$ & 2.12(0.11) &  $\cdots$ &  $\cdots$ & y &   &  y&  y &    &    &    \\ 
19.3 + 19.1 & 1.035 & 860(100) & 1.82(0.03) & 2.06(0.03) & 0.25(0.01) & 0.16(0.01) & y & y &  y&  y &  y &  y &  y \\ 
19.3 + 19.2 & $\cdots$  & 1400(150) &  $\cdots$ & 2.32(0.13) &  $\cdots$ &  $\cdots$ & y & y &  y&  y &  y &  y &  y \\ 
44.1 + 44.2 & 2.976 & 7580(140) & 1.27(0.03) & 1.46(0.03) & 0.38(0.42) & 0.22(0.24) & y & y &  y&  y &  y &    &    \\ 
44.1 + 44.3 & $\cdots$  & 4030(160) &  $\cdots$ & 1.82(0.04) &  $\cdots$ &  $\cdots$ & y & y &  y&  y &  y &    &    \\ 
59.3 + 59.1 & 1.875 & 1140(150) & 1.59(0.04) & 1.84(0.03) & 1.80(1.48) & 1.01(0.83) & y & y &  y&  y &  y &  y &  y \\ 
59.3 + 59.2 & $\cdots$  & 2960(180) &  $\cdots$ & 1.82(0.05) &  $\cdots$ &  $\cdots$ & y & y &  y&  y &  y &  y &  y \\ 
70.1 + 70.2 & 3.713 & 9910(180) & 1.25(0.03) & 1.04(0.04) & 0.09(0.09) & 0.05(0.05) & y &   &  y&  y &    &    &    \\ 
70.1 + 70.3 & $\cdots$  & 1730(150) &  $\cdots$ & 1.33(0.03) &  $\cdots$ &  $\cdots$ & y &   &  y&  y &    &    &  \\
\hline			
&&&MACSJ1149&&&  & &  &  &  & & \\ 
\hline
1.3 + 1.1 & 1.488 & 5950(200) & 1.33(0.04) & 2.28(0.05) & 2.37(0.35) & 1.50(0.14) & y & y &  y&  y &  y &  y &  y \\ 
1.3 + 1.2 & $\cdots$  & 6590(200) &  $\cdots$ & 1.74(0.09) &  $\cdots$ &  $\cdots$ & y & y &  y&  y &  y &  y &  y \\ 
2.1 + 2.2 & 1.894 & 840(210) & 1.66(0.06) & 4.69(0.54) & 0.05(0.02) & 0.03(0.01) & y & y &  y&  y &  y &  y &  y \\ 
2.1 + 2.3 & $\cdots$  & 840(200) &  $\cdots$ & 4.75(0.44) &  $\cdots$ &  $\cdots$ & y & y &  y&  y &  y &  y &  y \\ 
3.3 + 3.1 & 3.129 & 3150(300) & 1.61(0.09) & 2.22(0.08) & 5.82(1.28) & 3.47(0.75) & y & y &  y&  y &  y &  y &    \\ 
3.3 + 3.2 & $\cdots$  & 3560(250) &  $\cdots$ & 2.63(0.07) &  $\cdots$ &  $\cdots$ & y & y &  y&  y &  y &  y &    \\ 
4.3 + 4.1 & 2.949 & 9950(300) & 1.23(0.06) & 2.33(0.09) & 2.07(0.35) & 1.19(0.21) & y & y &  y&  y &  y &  y &    \\ 
4.3 + 4.2 & $\cdots$  & 10550(320) &  $\cdots$ & 1.99(0.06) &  $\cdots$ &  $\cdots$ & y & y &  y&  y &  y &  y &    \\ 
5.3 + 5.1 & 2.800 & 14230(300) & 1.30(0.05) & 2.82(0.08) & 1.30(0.37) & 0.80(0.23) & y & y &  y&  y &  y &  y &    \\ 
5.3 + 5.2 & $\cdots$  & 14290(300) &  $\cdots$ & 2.70(0.06) &  $\cdots$ &  $\cdots$ & y & y &  y&  y &  y &  y &    \\ 
13.3 + 13.1 & 1.240 & 920(80) & 1.82(0.03) & 2.67(0.07) & 0.13(0.04) & 0.08(0.02) & y & y &  y&  y &  y &  y &  y \\ 
13.3 + 13.2 & $\cdots$  & 990(60) &  $\cdots$ & 3.23(0.10) &  $\cdots$ &  $\cdots$ & y & y &  y&  y &  y &  y &  y \\ 
14.3 + 14.1 & 3.703 & 3340(440) & 1.68(0.15) & 4.17(0.27) & 0.008(0.002) & 0.005(0.001) & y &   &  y&  y &    &    &    \\ 
14.3 + 14.2 & $\cdots$  & 3330(450) &  $\cdots$ & 4.30(0.27) &  $\cdots$ &  $\cdots$ & y &   &  y&  y &    &    &    \\ 
29.3 + 29.1 & 3.214 & 4340(280) & 1.42(0.05) & 1.61(0.06) & 0.09(0.21) & 0.05(0.12) & y & y &  y&  y &  y &    &    \\ 
29.3 + 29.2 & $\cdots$  & 5870(280) &  $\cdots$ & 1.63(0.07) &  $\cdots$ &  $\cdots$ & y & y &  y&  y &  y &    &    \\ 
\hline
&&&MACSJ0416&&&  & &  &  &  & & \\ 
\hline
1.3 + 1.1 & 1.896 & 1270(120) & 1.13(0.03) & 1.72(0.04) & 2.02(0.46) & 1.16(0.26) & y & y &  y&  y &  y &  y &    \\ 
1.3 + 1.2 & $\cdots$  & 1600(120) &  $\cdots$ & 1.89(0.07) &  $\cdots$ &  $\cdots$ & y & y &  y&  y &  y &  y &    \\ 
2.3 + 2.1 & 1.893 & 2020(130) & 1.08(0.03) & 1.97(0.04) & 4.73(1.92) & 2.66(1.08) & y & y &  y&  y &  y &  y &    \\ 
2.3 + 2.2 & $\cdots$  & 2140(140) &  $\cdots$ & 2.04(0.11) &  $\cdots$ &  $\cdots$ & y & y &  y&  y &  y &  y &    \\ 
3.3 + 3.1 & 1.989 & 1180(120) & 1.06(0.04) & 1.16(0.04) & 0.52(0.05) & 0.33(0.04) & y & y &  y&  y &  y &  y &    \\ 
3.3 + 3.2 & $\cdots$  & 3870(90) &  $\cdots$ & 0.93(0.05) &  $\cdots$ &  $\cdots$ & y & y &  y&  y &  y &  y &    \\ 
4.3 + 4.1 & 1.989 & 1540(120) & 1.04(0.04) & 1.17(0.04) & 1.40(0.24) & 0.90(0.14) & y & y &  y&  y &  y &  y &    \\ 
4.3 + 4.2 & $\cdots$  & 4070(90) &  $\cdots$ & 0.87(0.06) &  $\cdots$ &  $\cdots$ & y & y &  y&  y &  y &  y &    \\ 
5.4 + 5.2 & 2.095 & 10820(180) & 0.75(0.03) & 2.53(0.07) & 0.64(0.38) & 0.39(0.21) & y & y &  y&  y &  y &  y &    \\ 
5.4 + 5.3 & $\cdots$  & 10850(180) &  $\cdots$ & 2.60(0.09) &  $\cdots$ &  $\cdots$ & y & y &  y&  y &  y &  y &    \\ 
7.3 + 7.1 & 2.086 & 6550(170) & 0.90(0.03) & 2.67(0.06) &  0.061(0.001) & 0.035(0.001)& y & y &  y&  y &  y &  y &    \\ 
7.3 + 7.2 & $\cdots$  & 6560(180) &  $\cdots$ & 2.55(0.08) &  $\cdots$ &  $\cdots$ & y & y &  y&  y &  y &  y &    \\ 
10.3 + 10.1 & 2.298 & 5940(160) & 0.86(0.04) & 1.77(0.05) & 1.01(0.21) & 0.60(0.12) & y & y &  y&  y &  y &  y &    \\ 
10.3 + 10.2 & $\cdots$  & 6440(180) &  $\cdots$ & 1.36(0.07) &  $\cdots$ &  $\cdots$ & y & y &  y&  y &  y &  y &    \\ 
11.3 + 11.1 & 1.006 & 1690(80) & 1.20(0.02) & 6.01(0.55) & 0.01(0.01) & 0.01(0.01) & y & y &  y&  y &  y &  y &  y \\ 
11.3 + 11.2 & $\cdots$  & 1690(80) &  $\cdots$ & 5.95(1.80) &  $\cdots$ &  $\cdots$ & y & y &  y&  y &  y &  y &  y \\ 
13.3 + 13.1 & 3.223 & 9110(190) & 0.70(0.04) & 1.88(0.05) & 1.79(0.87) & 1.14(0.61) & y &   &  y&  y &    &    &    \\ 
13.3 + 13.2 & $\cdots$  & 10540(210) &  $\cdots$ & 0.57(0.05) &  $\cdots$ &  $\cdots$ & y &   &  y&  y &    &    &    \\ 
14.1 + 14.2 & 1.637 & 3840(100) & 1.25(0.03) & 0.93(0.12) & 5.07(0.69) & 2.86(0.38) & y & y &  y&  y &  y &  y &    \\ 
14.1 + 14.3 & $\cdots$  & 1910(110) &  $\cdots$ & 1.33(0.05) &  $\cdots$ &  $\cdots$ & y & y &  y&  y &  y &  y &  y \\ 
15.3 + 15.1 & 2.336 & 7600(180) & 0.86(0.04) & 1.69(0.05) & 0.29(0.09) & 0.16(0.05) & y & y &  y&  y &  y &  y &    \\ 
15.3 + 15.2 & $\cdots$  & 8420(210) &  $\cdots$ & 0.62(0.06) &  $\cdots$ &  $\cdots$ & y & y &  y&  y &  y &  y &    \\ 
16.1 + 16.2 & 1.964 & 620(150) & 1.82(0.05) & 1.86(0.13) & 4.52(1.91) & 2.56(1.06) & y & y &  y&  y &  y &  y &  y \\ 
16.1 + 16.3 & $\cdots$  & 140(120) &  $\cdots$ & 1.32(0.05) &  $\cdots$ &  $\cdots$ & y & y &  y&  y &  y &  y &    \\ 
17.3 + 17.1 & 2.218 & 780(150) & 1.86(0.06) & 2.38(0.13) & 4.80(1.29) & 2.74(0.70) & y & y &  y&  y &  y &  y &    \\ 
17.3 + 17.2 & $\cdots$  & 790(160) &  $\cdots$ & 3.21(0.21) &  $\cdots$ &  $\cdots$ & y & y &  y&  y &  y &  y &    \\ 
23.1 + 23.2 & 2.091 & 4440(160) & 1.18(0.04) & 0.22(0.10) & 2.27(0.61) & 1.29(0.34) & y & y &  y&  y &  y &  y &    \\ 
23.1 + 23.3 & $\cdots$  & 440(140) &  $\cdots$ & 1.05(0.03) &  $\cdots$ &  $\cdots$ & y & y &  y&  y &  y &  y &    \\ 
27.3 + 27.1 & 2.107 & 650(110) & 1.50(0.04) & 1.84(0.03) & 0.33(0.37) & 0.20(0.20) & y & y &  y&  y &  y &  y &    \\ 
27.3 + 27.2 & $\cdots$  & 680(100) &  $\cdots$ & 2.48(0.11) &  $\cdots$ &  $\cdots$ & y & y &  y&  y &  y &  y &    \\ 
29.1 + 29.2 & 2.285 & 5420(160) & 0.95(0.03) & 3.07(3.76) & 5.19(1.50) & 2.99(0.83) & y & y &  y&  y &  y &  y &    \\ 
29.1 + 29.3 & $\cdots$  & 2510(160) &  $\cdots$ & 1.20(0.04) &  $\cdots$ &  $\cdots$ & y & y &  y&  y &  y &  y &    \\ 
35.1 + 35.2 & 3.492 & 6150(200) & 0.97(0.05) & 0.85(0.09) & 0.47(0.02) & 0.27(0.03) & y &   &  y&  y &    &    &    \\ 
35.1 + 35.3 & $\cdots$  & 3450(170) &  $\cdots$ & 1.40(0.04) &  $\cdots$ &  $\cdots$ & y &   &  y&  y &    &    &    \\ 
38.3 + 38.1 & 3.440 & 9160(240) & 1.05(0.04) & 1.44(0.06) & 0.25(0.03) & 0.14(0.02) & y &   &  y&  y &    &    &    \\ 
38.3 + 38.2 & $\cdots$  & 9650(310) &  $\cdots$ & 0.84(0.14) &  $\cdots$ &  $\cdots$ & y &   &  y&  y &    &    &    \\ 
44.3 + 44.1 & 3.290 & 1120(160) & 1.17(0.04) & 1.29(0.05) & 1.21(1.06) & 0.80(0.70) & y &   &  y&  y &    &    &    \\ 
44.3 + 44.2 & $\cdots$  & 240(150) &  $\cdots$ & 1.84(0.07) &  $\cdots$ &  $\cdots$ & y &   &  y&  y &    &    &    \\ 
47.3 + 47.1 & 3.253 & 10030(220) & 0.78(0.04) & 2.05(0.07) & 0.33(0.02) & 0.208(0.001) & y &   &  y&  y &    &    &    \\ 
47.3 + 47.2 & $\cdots$  & 10370(260) &  $\cdots$ & 1.40(0.07) &  $\cdots$ &  $\cdots$ & y &   &  y&  y &    &    &    \\ 
\hline
    \multicolumn{13}{c}{MACSJ0416} \\ 
			\hline
		49.3 + 49.1 & 3.871 & 6110(190) & 1.06(0.04) & 0.49(0.07) & 0.53(0.02) & 0.29(0.01) & y &   &   &  y &    &    &    \\ 
		49.3 + 49.2 & $\cdots$  & 1160(200) &  $\cdots$ & 0.86(0.04) &  $\cdots$ &  $\cdots$ & y &   &   &  y &    &    &    \\ 
		55.2 + 55.1 & 3.292 & 14810(220) & 0.66(0.03) & 1.61(0.04) & 0.23(0.12) & 0.15(0.09) & y &   &  y&  y &    &    &    \\ 
		86.3 + 86.1 & 3.292 & 3130(180) & 0.92(0.03) & 1.35(0.04) & 0.51(0.36) & 0.32(0.20) & y &   &  y&  y &    &    &    \\ 
		86.3 + 86.2 & $\cdots$  & 6930(200) &  $\cdots$ & 0.79(0.11) &  $\cdots$ &  $\cdots$ & y &   &  y&  y &    &    &    \\ 
			\hline
    \multicolumn{13}{c}{MACSJ0717}\\ 
			\hline
1.5 + 1.1 & 2.963 & 52870(1220) & 1.79(0.09) & 5.24(0.56) & 0.42(0.04) & 0.24(0.02) & y & y &  y&  y &  y &  y &    \\ 
1.5 + 1.2 & $\cdots$  & 52840(1200) &  $\cdots$ & 4.40(0.16) &  $\cdots$ &  $\cdots$ & y & y &  y&  y &  y &  y &    \\ 
1.5 + 1.3 & $\cdots$  & 52940(1260) &  $\cdots$ & 4.17(0.20) &  $\cdots$ &  $\cdots$ & y & y &  y&  y &  y &  y &    \\ 
1.5 + 1.4 & $\cdots$  & 51460(1420) &  $\cdots$ & 2.41(0.17) &  $\cdots$ &  $\cdots$ & y & y &  y&  y &  y &  y &    \\ 
3.3 + 3.1 & 1.855 & 17210(1610) & 1.32(0.04) & 2.32(0.09) & 0.34(0.10) & 0.19(0.05) & y & y &  y&  y &  y &  y &    \\ 
3.3 + 3.2 & $\cdots$  & 16510(1460) &  $\cdots$ & 2.98(0.06) &  $\cdots$ &  $\cdots$ & y & y &  y&  y &  y &  y &    \\ 
4.3 + 4.1 & 1.855 & 18020(400) & 1.52(0.05) & 2.91(0.09) & 0.79(0.40) & 0.44(0.22) & y & y &  y&  y &  y &  y &  y \\ 
4.3 + 4.2 & $\cdots$  & 15250(430) &  $\cdots$ & 1.68(0.06) &  $\cdots$ &  $\cdots$ & y & y &  y&  y &  y &  y &  y \\ 
12.2 + 12.1 & 1.710 & 7020(340) & 1.46(0.05) & 1.74(0.08) & 0.99(0.19) & 0.57(0.10) & y & y &  y&  y &  y &  y &  y \\ 
12.2 + 12.3 &$\cdots$& 1580(520) &$\cdots$ & 2.05(0.04) & $\cdots$ & $\cdots$ & y & y &  y&  y &  y &  y &  y \\ 
13.3 + 13.1 & 2.547 & 22600(820) & 1.73(0.04) & 1.96(0.06) & 12.74(9.23) & 7.17(5.14) & y & y &  y&  y &  y &  y &    \\ 
13.3 + 13.2 & $\cdots$  & 7560(820) &  $\cdots$ & 1.42(0.05) &  $\cdots$ &  $\cdots$ & y & y &  y&  y &  y &  y &    \\ 
14.2 + 14.1 & 1.855 & 14630(530) & 1.45(0.03) & 2.01(0.28) & 1.93(0.88) & 1.09(0.49) & y & y &  y&  y &  y &  y &  y \\ 
14.2 + 14.3 & $\cdots$ & 10710(940) & $\cdots$ & 2.58(0.08) & $\cdots$& $\cdots$ & y & y &  y&  y &  y &  y &  y 
\enddata
\tablenotetext{a}{Population of intrinsically faint CC SNe ($\rm M_B >-15$) which could be contributing importantly in the CC SN rates. See e.g. \citet{2015ApJ...813...93S} for a discussion.}

\end{deluxetable}
\end{longrotatetable}


\end{document}